\pdfoutput=1
\documentclass[aps,twocolumn,prb,groupedaddress]{revtex4}
\usepackage{graphicx}
\usepackage{amsmath}
\usepackage{amssymb}
\usepackage[T1]{fontenc}
\usepackage{verbatim}
\usepackage{epstopdf}

\newcommand{\be}{\begin{equation}}
\newcommand{\ee}{\end{equation}}
\newcommand{\bd}{\begin{displaymath}}
\newcommand{\ed}{\end{displaymath}}
\newcommand{\bea}{\begin{eqnarray}}
\newcommand{\eea}{\end{eqnarray}}
\newcommand{\Eq}[1]{Eq.(\ref{#1})}% \Eq{abc}
\newcommand{\Fig}[1]{Fig.\,\ref{#1}}% \Fig{abc}
\newcommand{\Sec}[1]{Sec.\,\ref{#1}}% \Fig{abc}
\newcommand{\Onlinecite}[1]{Ref.\,\onlinecite{#1}} % \Onlinecite{abc}
\newcommand{\Tab}[1]{Table \,\ref{#1}}% \Tab{abc}

\newcommand{\br}{\mathbf{r}}
\newcommand{\lc}{\lambda_{\rm c}}
\newcommand{\wc}{\omega_{\rm c}}
\newcommand{\Sf}{S_{\rm m}}

\begin{document}
\textheight 25cm %\preprint{}
\sloppy
\title{Structure and zero-dimensional polariton spectrum
\\of natural defects in GaAs/AlAs microcavities}
\author{Joanna M Zajac}
\email[Electronic address:]{ZajacJM@cardiff.ac.uk}
\author{Wolfgang Langbein}
\affiliation{School of Physics and Astronomy, Cardiff University,
The Parade, Cardiff CF24 3AA, United Kingdom}

\date{\today}
\begin{abstract}
We present a correlative study of structural and optical properties of natural defects in
planar semiconductor microcavities grown by molecular beam epitaxy, which are showing a
localized polariton spectrum as reported in Zajac {\em et al.}, Phys. Rev. B {\bf 85}, 165309
(2012). The three-dimensional spatial structure of the defects was studied using combined
focussed ion beam (FIB) and scanning electron microscopy (SEM). We find that the defects
originate from a local increase of a GaAs layer thickness. Modulation heights of up to 140\,nm
for oval defects and 90\,nm for round defects are found, while the lateral extension is about
2$\mu$m for oval and 4$\mu$m for round defects. The GaAs thickness increase is attributed to
Ga droplets deposited during growth due to Ga cell spitting. Following the droplet deposition,
the thickness modulation expands laterally while reducing its height, yielding oval to round
mounds of the interfaces and the surface. With increasing growth temperature, the ellipticity
of the mounds is decreasing and their size is increasing. This suggests that the expansion is
related to the surface mobility of Ga, which with increasing temperature is increasing and
reducing its anisotropy between the $[110]$ and $[1\bar{1}0]$ crystallographic directions.
Comprehensive data consisting of surface profiles of defects measured using differential
interference contrast (DIC) microscopy, volume information obtained using FIB/SEM, and
characterization of the resulting confined polariton spectrum are presented.
\end{abstract}
\maketitle

\section{Introduction}
Fundamental physics was demonstrated in planar semiconductor microcavities over last 10 years
including Bose-Einstein condensation of exciton-polaritons\,\cite{KasprzakNature06}, formation
of vortices\,\cite{LagoudakisNatPhys08} and superfluidity\,\cite{AmoNature09}. To understand
these two-dimensional inhomogeneous non-equilibrium systems,  it is important to understand
and harness spatial disorder in these structures. A significant contribution to polariton
disorder in molecular beam epitaxy (MBE) grown microcavities is photonic disorder including a
cross-hatched dislocation pattern\,\cite{ZajacAPL12,AbbarchiPRB12} and
point-like-defects\,(PDs)\cite{ZajacPRB12}. Such disorder creates a potential landscape for
the in-plane motion of polaritons, which results in inhomogeneous broadening, enhanced
backscattering \cite{LangbeinPRL02a,GurioliPRL05}, and creation of localized polariton
condensates and polariton vortices\,\cite{LagoudakisNatPhys08,KrizhanovskiiPRB09}.

In this work, we present a correlative study of the structural and optical properties of
natural defects in planar semiconductor GaAs/AlAs microcavities grown by MBE. The paper is
organized as follows: in \Sec{sec:defects} we review the literature on defects in MBE grown
GaAs structures, in \Sec{sec:samples} we discuss the samples and experimental methods used, in
\Sec{sec:results} we present the experimental results, followed by a discussion of the defect
formation in \Sec{sec:disc} and conclusions in \Sec{sec:sum}.

\section{Point-like defects in {GaAs} heterostructures}
\label{sec:defects} Point-like-defects observed in MBE-grown GaAs based heterostructures are
classified into several types, with the most common being oval defects. It was reported that
oval defects originate from an excess of Ga, Ga droplets, or surface
contamination\cite{ChandJCrystGrowth90,FujiwaraJCrystGrowth87}, or low growth
temperatures\cite{OrmeMRSSP94,KreutzerUltramicroscopy99}. Oval defects were extensively
studied in the 1990s as they were responsible for the failure of electronic devices such as
field-effect transistors.\cite{ChandJCrystGrowth90} The defects have typical sizes in the
order of several $\mu$m and a roughly 3:1 aspect ratio along the $[1\bar{1}0]$:$[110]$ crystal
directions.\cite{FujiwaraJCrystGrowth87} Their height on the surface were found to be several
tens of nanometers. Another type of defect observed in this work were round defects having
similar diameters and heights as oval defects. They were attributed\cite{KawadaJCG93} to Ga
oxide or effusion cell spitting. We found that the defects investigated in our work originate
from a GaAs thickness modulation, which we attribute to Ga droplets with sizes in the order of
100\,nm emitted by the Ga source during growth. The droplet formation was previously
ascribed\cite{ChandJCrystGrowth90} to an inhomogeneous temperature distribution in the Ga
crucible of the effusion cell. Specifically, Ga cools near the orifice of the crucible and,
since it does not wet the pyrolytic Boron Nitride (PBN) crucible surface, forms droplets which
can fall back into the liquid Ga, causing a spatter of smaller Ga droplets. Recommended
methods to reduce this mechanism include 1) use solid instead of liquid Ga; 2) modification of
the orifice geometry of the Ga cell to inhibit condensed Ga droplets entering into the Ga
source; 3) creating a positive axial temperature gradient toward the orifice to prevent
condensation of Ga; 4) treating the crucible with Al, which forms a AlN layer which Ga is
wetting, suppressing the formation of droplets. Another mechanism for formation of oval
defects is suggested in \Onlinecite{BrunemeierJVacTechnolB91} and referred to as Ga source
spitting. During heating of the Ga source up to $1200^\circ$C, within the range of Ga
evaporation \cite{Herman89MBE}, explosions in the Ga liquid were observed, which resulted in
Gallium droplets deposition on the MBE chamber walls. It was speculated that these explosions
were due to Ga$_{2}$O$_{3}$ shells encapsulating Ga and creating an effusion barrier. Another
possible mechanism\cite{ClarkePrivateCommunication} is that particulates released from the
walls of the MBE chamber are entering the molten Ga in the crucible causing a turbulent
reaction. Summarizing, a number of mechanisms for the Ga nano-droplet formation have been
suggested, and the mechanism dominant in a given growth is not obvious.

\section{Samples and Experimental methods}
\label{sec:samples} %MC1=VN1342 %MC2=VN1968 \label{sec:setup}
In this work we investigated two microcavity samples, MC1 and MC2, grown in a VG Semicon V90
MBE machine with a hot-lip Veeco 'SUMO' cell as Ga source, with structures given in
\Tab{tab:Samples}. Sample MC1 was studied in \Onlinecite{ZajacPRB12}.
\begin{table}
\begin{center}
\renewcommand{\arraystretch}{1.3}
    \begin{tabular*}{\columnwidth}{@{\extracolsep{\fill}}l|r|c c }
    \hline
    \multicolumn{2}{l|}{Sample} & MC1 & MC2  \\
    \hline
    \multicolumn{2}{l|}{cavity length} & 1$\lambda_{c}$ & 2$\lambda_{c}$ \\
    \multicolumn{2}{l|}{DBR periods top(bottom)} & 24(27) & 23(26) \\
    \hline
    growth  & DBR AlAs   & 715 & 590 \\
    temperature & DBR GaAs    & 660 & 590 \\
    ($^\circ $C)& cavity GaAs & 630 & 590 \\
    \hline
    \end{tabular*}
    \caption{Parameters of samples MC1 and MC2.}
    \label{tab:Samples}
 \end{center}
 \end{table}
During the growth of MC1, the wafer temperature was ramped up to $715\,^\circ $C for the AlAs
Bragg layers and down to $660^\circ $C for GaAs Bragg layers, while the cavity layer was grown
at $630\,^\circ$C. During the growth of MC2 instead, the growth temperature was $590\,^\circ
$C for all layers. The two samples show a significantly different aspect ratio of defects on
their surface. In MC1 they are essentially round (see \Fig{fig:D111015018fib}), while in MC2
they have an aspect ratio between 3:1 and 2:1 along the $[1\bar{1}0]$ : $[110]$ direction as
shown in \Fig{fig:D110906006fib}. At a temperature of T=80\,K, the cavity mode energy in the
center of the wafer of MC1 (MC2) is at $\hbar\wc=1.480(1.431)$\,eV, respectively, while the
bulk GaAs exciton resonance of the cavity layer is at 1.508\,eV.

\begin{figure}[t]
\includegraphics[width=\columnwidth]{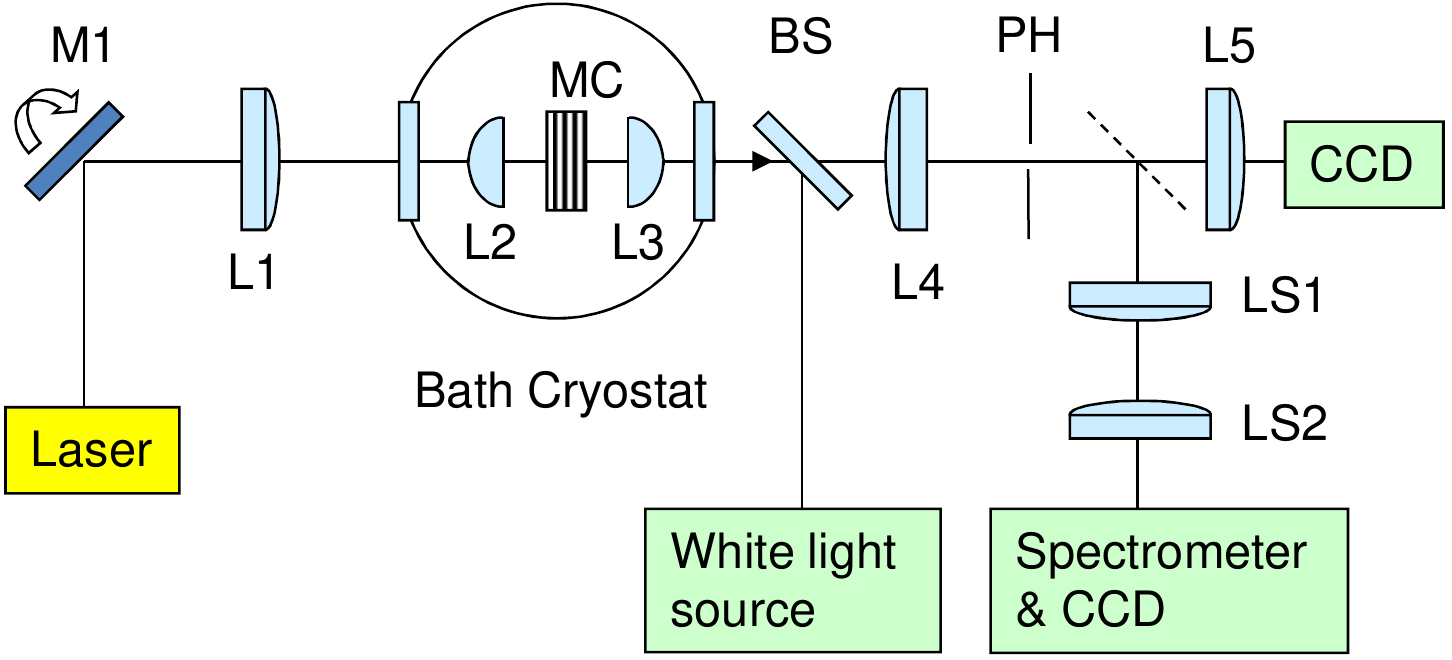}
\caption{Sketch of the optical imaging spectroscopy setup used to measure the localized
polariton states. M1: Gimbal mounted mirror, L1-L5: Lenses, MC: Microcavity sample, LS1,LS2:
movable lenses for imaging, dashed lines: removable mirrors, BS: Beam-splitter.}
\label{fig:setup}
\end{figure}

The low temperature optical measurements were performed using the optical setup sketched in
\Fig{fig:setup}. The samples were mounted strain-free on a mechanical translation stage moving
along the sample surface $(x,y)$ in a bath cryostat at $T=80$\,K in nitrogen gas at
100-300\,mbar. Two aspheric lenses of 8\,mm focal length and 0.5 numerical aperture (NA) were
mounted at the opposing faces of the sample inside the cryostat to focus the excitation and
collimate the emission, respectively, providing a diffraction limited resolution of $1\,\mu$m.
The axial positions of both lenses were adjustable at low temperatures to control the focus of
excitation and detection. The excitation was provided by a mode-locked Ti:Sapphire laser
(Coherent Mira) delivering 100\,fs pulses at 76\,MHz repetition rate and a spectral width of
approximately 20\,meV. The transmission of the samples excited from the substrate side and
detected from the epi-side was imaged onto the input slit of an imaging spectrometer with a
spectral resolution of $15\,\mu$eV. Scanning of the sample image across the spectrometer input
slit while keeping the directional image on the spectrometer grating fixed for two-dimensional
hyperspectral imaging was achieved by moving the lenses LS1 and LS2
appropriately\cite{LangbeinRNC10}.

Spatial height profiles of defects on the sample surface were measured with differential
interference contrast (DIC) microscopy using an Olympus BX-50 microscope with a 20x 0.5NA
objective. DIC images in green light (wavelength range 525-565\,nm) were taken by a Canon EOS
500D camera with an array of 4752 x 3168 pixels of $4.8\,\mu$m size in the intermediate image
plane. The resulting image resolution was $650$\,nm in the plane of the sample, and about
$2$\,nm in the plane perpendicular to the sample surface using quantitative DIC. Details on
the procedure used to extract height profiles of defects using DIC microscopy are given in the
Appendix.

\begin{figure}
\includegraphics[width=0.8\columnwidth]{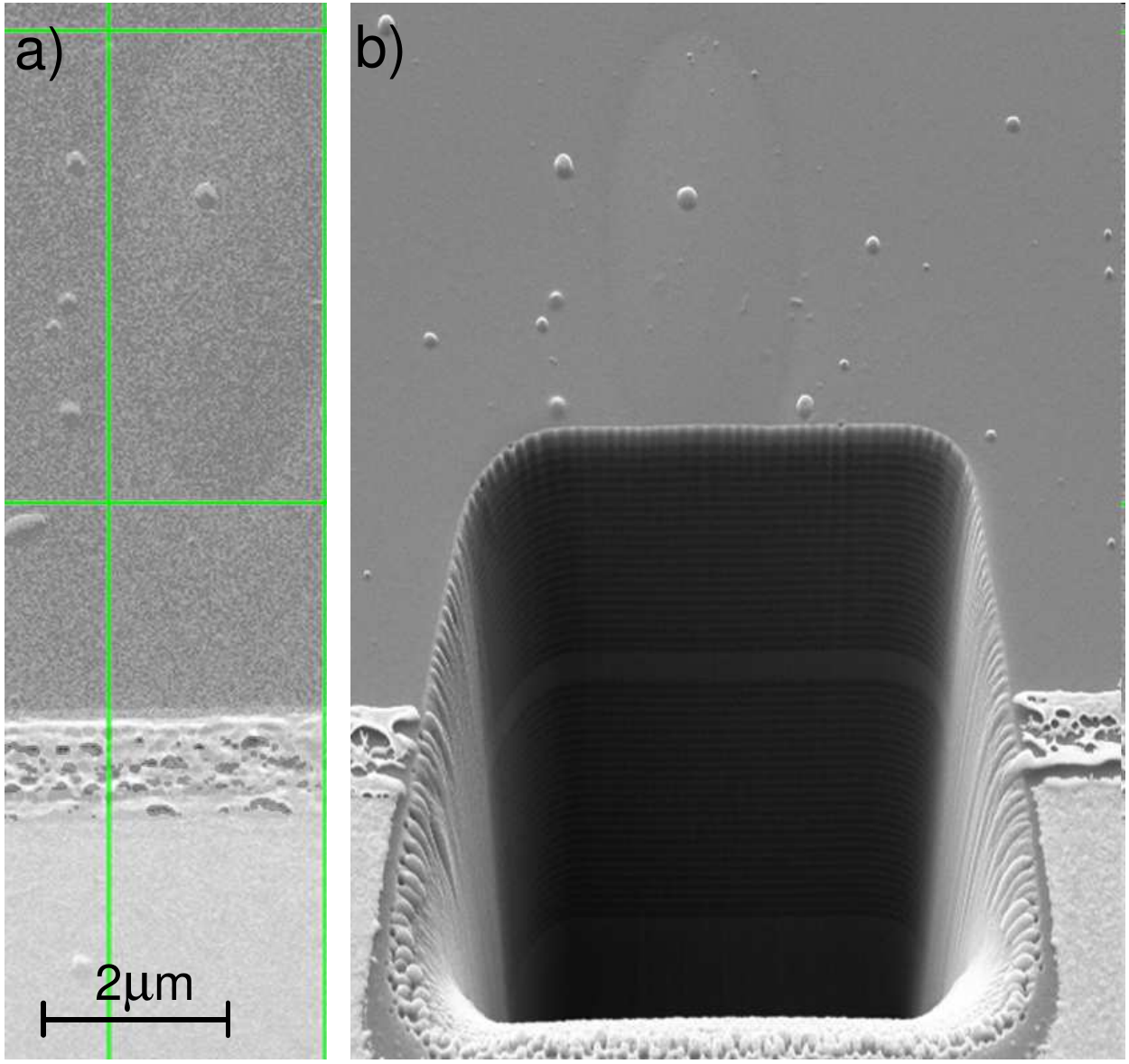}
\caption{SEM images of oval defect PD3. a) prior to milling, green lines indicate defect
extension on the surface. b) after milling the well exposing the epilayer cross-section at its
side wall. On the lower part of the images, the edge of the alignment photomask is visible.}
\label{fig:FIB}
\end{figure}
\begin{figure*}[]
\includegraphics[width=\textwidth]{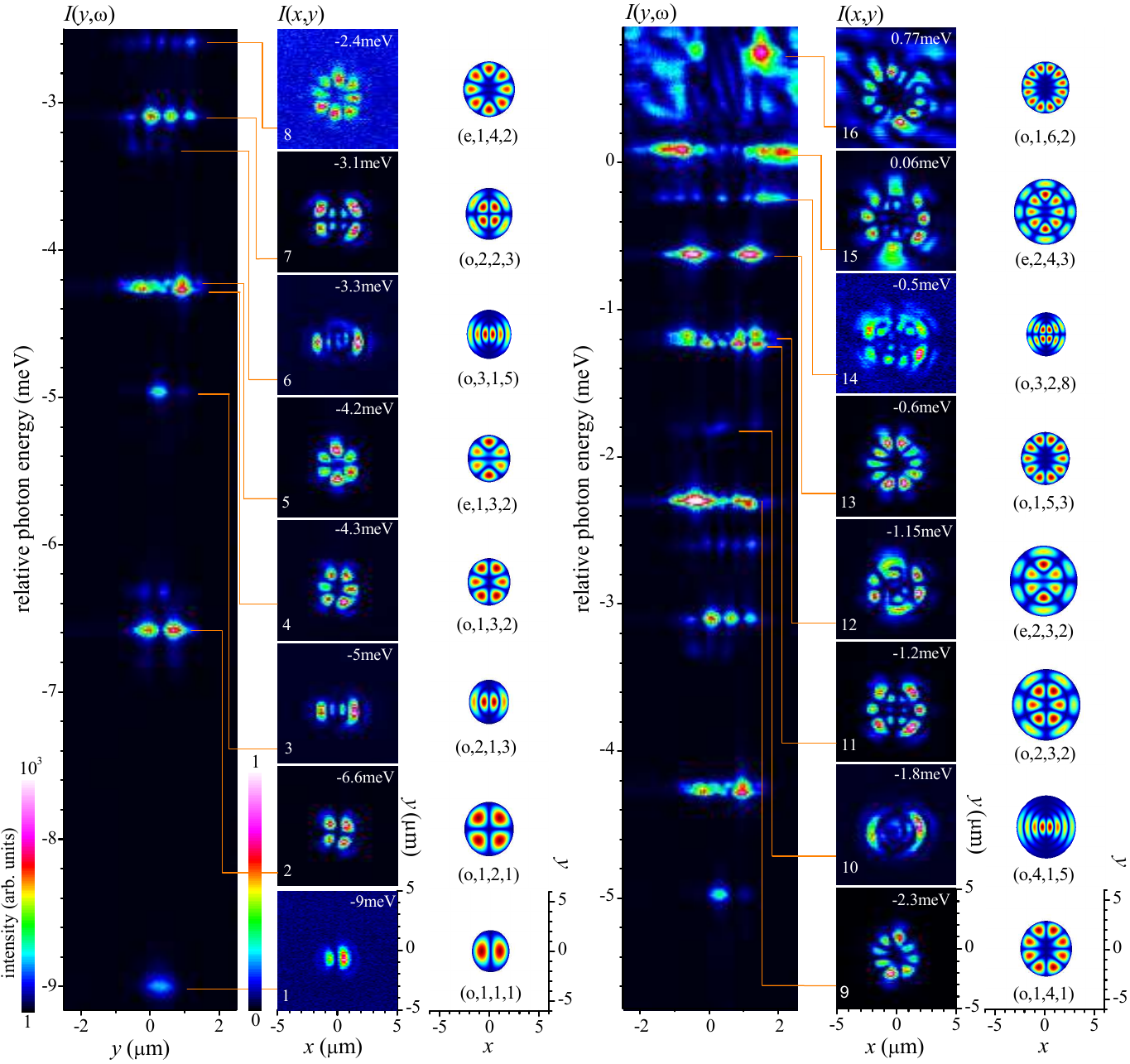}
\caption{Hyperspectral imaging of polariton states bound to PD1 in MC1, measuring the
spatially and spectrally resolved transmission intensity $I(x,y,\omega)$. First and fourth
column: $I(0,y,\omega)$. The energy is shown relative to the polariton band edge at
$\hbar\omega_{\rm c}=1.4789$\,eV. Intensity on a logarithmic color scale as indicated. Second
and fifth column: Real space intensity maps of individual states $I(x,y,\omega_l)$. The state
number $l$ and the relative energy $\hbar(\omega_l-\omega_{\rm c})$ are given, and the orange
lines indicate the related peaks in $I(0,y,\omega)$. Third and sixth column: Real space
distributions $|\Psi_{n,m,q}|^2$ of corresponding Mathieu functions, labeled by their parity
('e': even, 'o': odd), radial ($n$) and angular ($m$) order, and parameter $q$ (see
\Eq{eq:mathieuPhi} and \Eq{eq:mathieuR}). Color scale as for measured data indicated in the
first column.} \label{fig:optical111015018}.
\end{figure*}
To investigate the sample structure below the surface, we used a Carl Zeiss XB1540 Cross-Beam
focussed-ion-beam (FIB) microscope which combines an ion-beam milling/imaging column with
field-emission scanning electron microscope (FESEM) \cite{Giannuzzi05FIB}, providing an
imaging resolution of about $50$\,nm for the samples studied. The internal epi-layer structure
was exposed by FIB milling. Smooth cross-sections were obtained using a two stage milling
procedure. In the first step, a high beam current of $2\,$nA, was used, resulting in fast
milling but leaving a rough and inhomogeneous interface due to sputtering and redeposition of
material. In the second step, the surface was polished with a lower beam current of $200\,$pA
removing a layer of about $500\,$nm per cut resulting in a negligible surface roughness. In
the next steps, layers with a thickness of $1\,\mu$m or less were removed using the low beam
current. Different stages of this milling procedure are shown in \Fig{fig:FIB}. Before
milling, the oval defect (PD3) is seen in SEM (\Fig{fig:FIB}a), with the vertical image scale
corrected for the viewing angle of $36^\circ$ to the sample surface. A rectangular well of
about $14\,\mu$m width and $7\,\mu$m depth is milled into the surface to one side of the
defect (\Fig{fig:FIB}b), exposing a cross-section through the epi-layers to be measured at its
side walls. After imaging the exposed cross-section with SEM, the subsequent cross-section at
a controlled distance further into the structure is milled. Steps sizes of about $1\,\mu$m
were used at the outskirts of the defect, reducing down to $100\,$nm at its center to resolve
the defect source. The cross-sections were $6-7\,\mu$m deep to expose the complete epitaxial
structure, and their width was adjusted to the lateral extension of the defect observed at the
surface. In order to mark defects on the sample surface for the correlative studies, a gold
alignment mask was fabricated on the surface by photolithography. The mask consisted of a grid
of $400\,\mu$m$\,\times\,400\,\mu$m squares with column and row indexing. Considering the
defect density in the order of $10^{3}$/cm$^{2}$, this mask allows to trace individual defects
through the different measurement techniques used in the present investigation.
\section{Results}
\label{sec:results}
\subsection{Sample MC1}
\label{sec:MC1}
The localized polariton states in the round defects of sample MC1 were examined previously in
\Onlinecite{ZajacPRB12}. Here we report on the correlation between the states and the
three-dimensional structure of these defects using two defects referred to as PD1 and PD2 as
examples. The localized polariton states of PD1 are observed in the hyperspectral transmission
images shown in \Fig{fig:optical111015018}. The defect shows 16 localized modes down to
$-9$\,meV below the polariton band edge. It can be noted that the lowest observed stated is of
$p_x$-type symmetry. We expect to have 2 states with lower energies, another $p_y$-state and
an $s$-state. These were not recorded.
\begin{figure}
\includegraphics[width=\columnwidth]{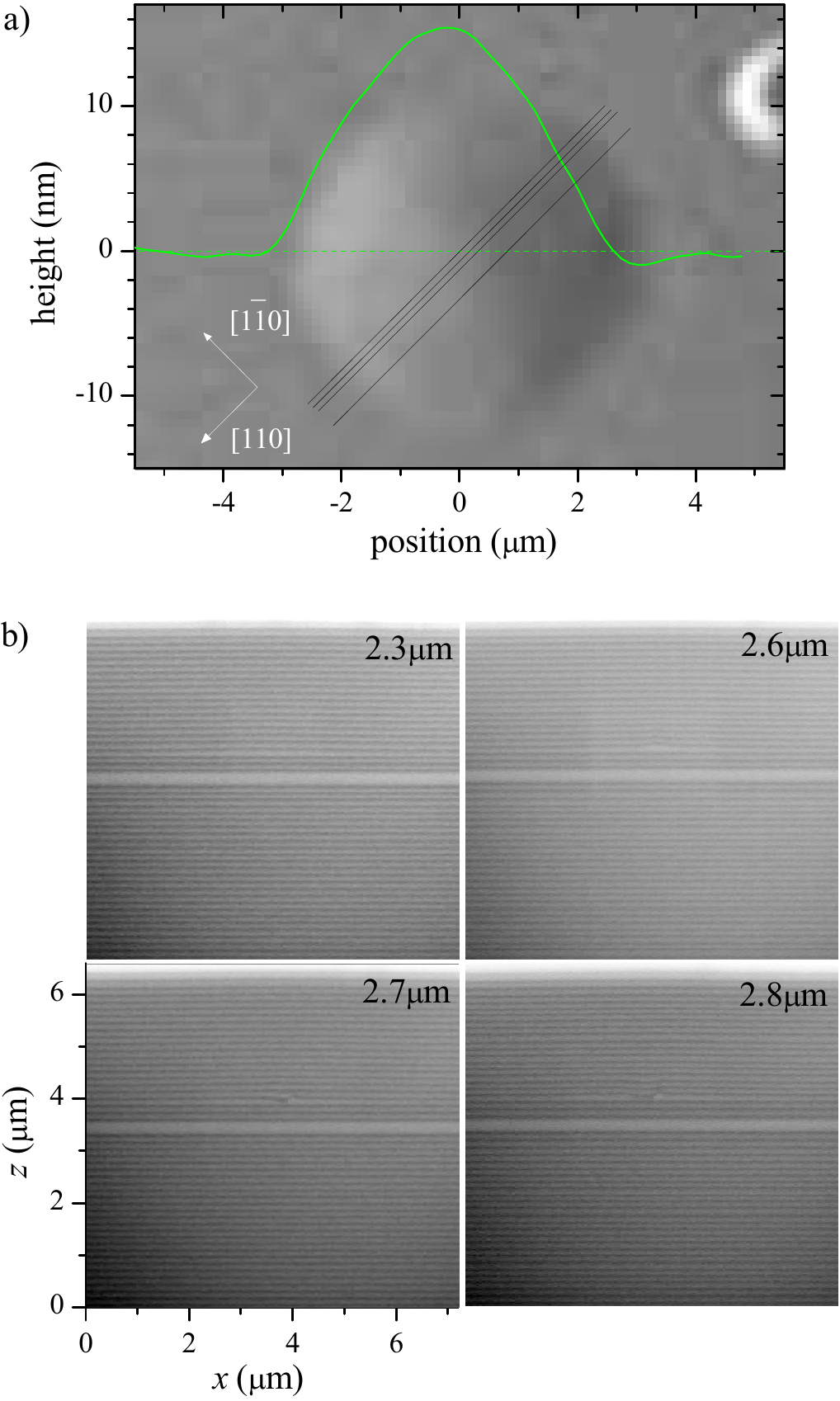}
\caption{Structural characterization of PD1. a) DIC image of the sample surface (linear grey
scale) and resulting height profile across the defect center (green line). b) SEM images of
cross-sections through the epitaxial structure, taken along the black lines indicated in a).
The relative distances from the edge of the defect are indicated.} \label{fig:D111015018fib}
\end{figure}

To qualitatively understand the states bound to this defect, we compare them with solutions of the
two-dimensional time-dependent wave equation
$\left(\partial_x^2+\partial_y^2-\partial_t^2/v^2\right)\Psi=0$ with the velocity $v$ for
elliptical boundary conditions. Using elliptical coordinates $x = c\cosh{(\xi)}\cos{(\eta)}$, $y =
c\sinh{(\xi)}\sin{(\eta)}$ with the focus distance $c$, and the ansatz
$\Psi(x,y,t)=R(\xi)\Phi(\eta)e^{i\omega t}$ results in the ordinary and modified Mathieu equations
for the angular part $\Phi(\xi)$ and radial part $R(\xi)$, respectively:\cite{Abramowitz65}
\bea
\left(\partial_\eta^2 +  \lambda -2q\cos{(2\eta)}\right)\Phi = 0 \label{eq:mathieuPhi}\\
\left(\partial_\xi^2  + \lambda -2q\cosh{(2\eta)}\right)R = 0 \label{eq:mathieuR} \eea
where $q=(\omega c/2v)^2$ is the square of the normalized frequency. The solutions of $\Phi$
and $R$ are angular and radial Mathieu functions. For a given $q$, the periodic boundary
condition of the angular part $\Phi$ in \Eq{eq:mathieuPhi} determines a series of
$\lambda_m(q)$ with ascending number of nodes $m=0,1,2,...$, each of which except $m=0$ is a
doublet, having an odd (o) or even (e) symmetry for inversion of $y$. The elliptical boundary
is given by a unique $c$ and $\xi_0$, at which the boundary condition, for example
$R(\xi_0)=0$, holds. This condition and \Eq{eq:mathieuR} using $\lambda=\lambda_m(q)$
determines the mode frequencies $q_{n,m}$ corresponding to modes with ascending number of
nodes $n=0,1,2,..$ in radial direction. Analytic expressions for the solutions are given in
\Onlinecite{McLachlan47}.

Since the polaritons show a quadratic in-plane dispersion for small wavevectors, their in-plane
motion is well described by the Schr{\"o}dinger equation.\cite{KaitouniPRB06} By modifying the
definition of $q$, the above solutions for the Helmholtz equation are also solving the
Schr{\"o}dinger equation for elliptical boundary conditions, such as the elliptical quantum well
with infinite barriers.\cite{WaalkensAnnPhys97}

Another family of analytic solutions of the Schr{\"o}dinger equation with elliptical symmetry
are given by the Hermite-Gaussian modes of an anisotropic two-dimensional harmonic oscillator.
However, we found that they do not describe the observed distributions well, since, as we will
see later, the confining potential of PD1 is not similar to a parabolic potential but rather
to an elliptical well.

The solutions\cite{Scilab} $|\Psi_{n,m,q}|^2$ with the mode orders $n$, $m$ assigned to the
measured states of intensities $|\psi_l|^2$ are shown in \Fig{fig:optical111015018}, yielding a
qualitative agreement. For each energy state, $q$ was adjusted in order to reproduce the
experimental patterns.

%It can be noticed that the quality of FIB data for this samples is worse that for a previous
%sample, the reason is twofold, firstly, the aspect ration of the defect's size is smaller and seems
% without deposition of metallic film what resulted in charging effects.

The structural characterization of PD1 by DIC and FIB/SEM is given in \Fig{fig:D111015018fib}.
The DIC data shows a diameter of the defect on the surface of about $6\,\mu$m, similar to the
extension of the spatially resolved transmission from this defect, and a height up to about
$15$\,nm. The FIB/SEM cross-sections show the origin of the defect - a thickened GaAs layer
with a center depression in the third DBR period above the cavity, extending over $4\,\mu$m in
$x$, and adding about $60$\,nm in thickness in the center, visible in the $y=4.5\,\mu$m
cross-section. Further discussion of the defect growth dynamics will be given later.

\begin{figure}
\includegraphics[width=\columnwidth]{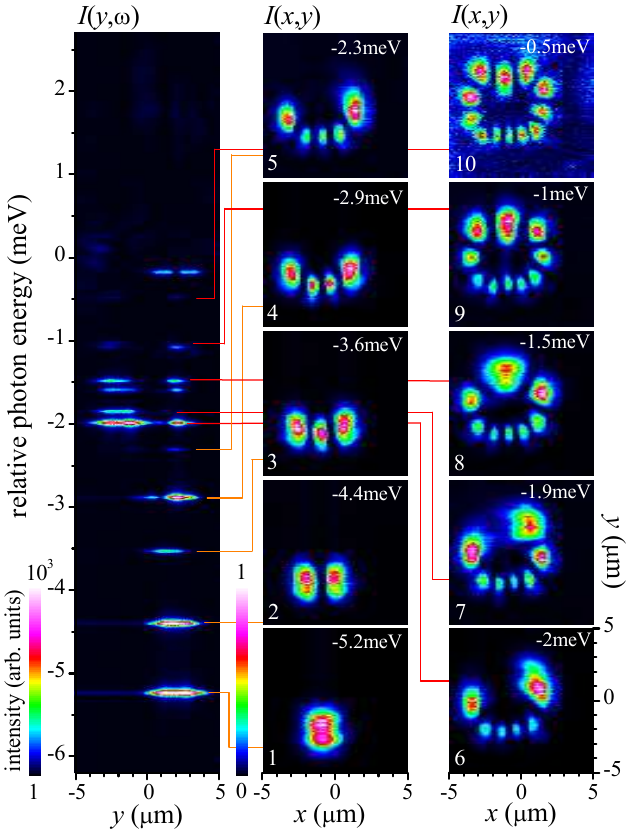}
\caption{Hyperspectral imaging of polariton states bound to PD2 in MC1 with $\hbar\wc=1.4787$\,eV,
detailed description as for \Fig{fig:optical111015018}.} \label{fig:D111015012}
\end{figure}

We now move to the second defect PD2 on MC1, for which the hyperspectral transmission images are
shown in \Fig{fig:D111015012}. The states are arranged along a ring of about $6\,\mu$m diameter,
with the lowest state localized at small $y$, and higher states gradually extending along the ring,
as in a one-dimensional harmonic oscillator, until the whole ring is filled. The underlying
potential for the polaritons appears to be a ring-shaped well, with a depth decreasing with
increasing $y$. The analysis of the potential from the states shown in \Sec{sec:potential} confirms
this interpretation.

\begin{figure}
\includegraphics[width=\columnwidth]{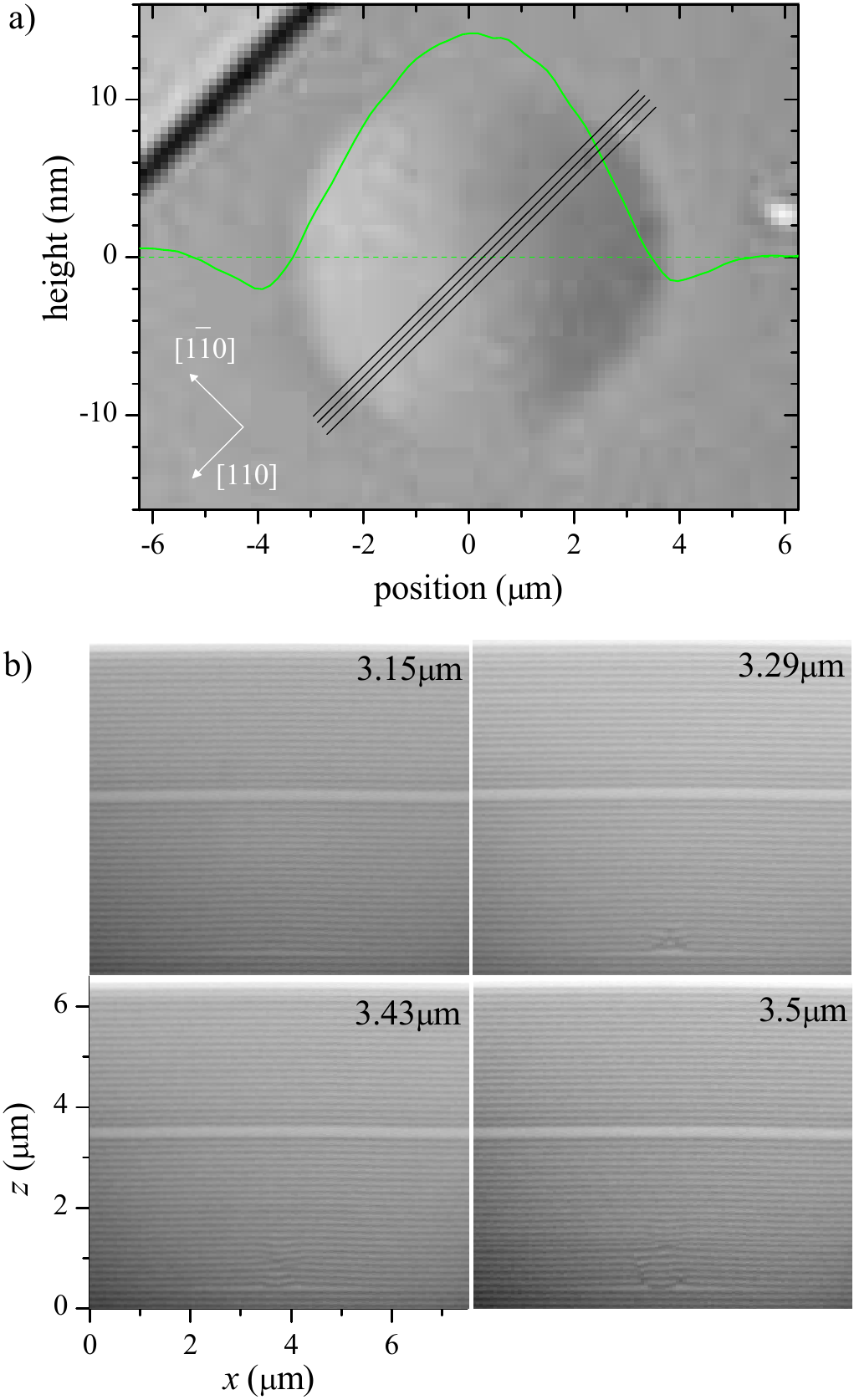}
\caption{Structural characterization of PD\,2, detailed description as for
\Fig{fig:D111015018fib}.}
%22 top and 26 bottom DBRs is shown.}
\label{fig:D111015012fib}
\end{figure}

\begin{figure}
\includegraphics[width=\columnwidth]{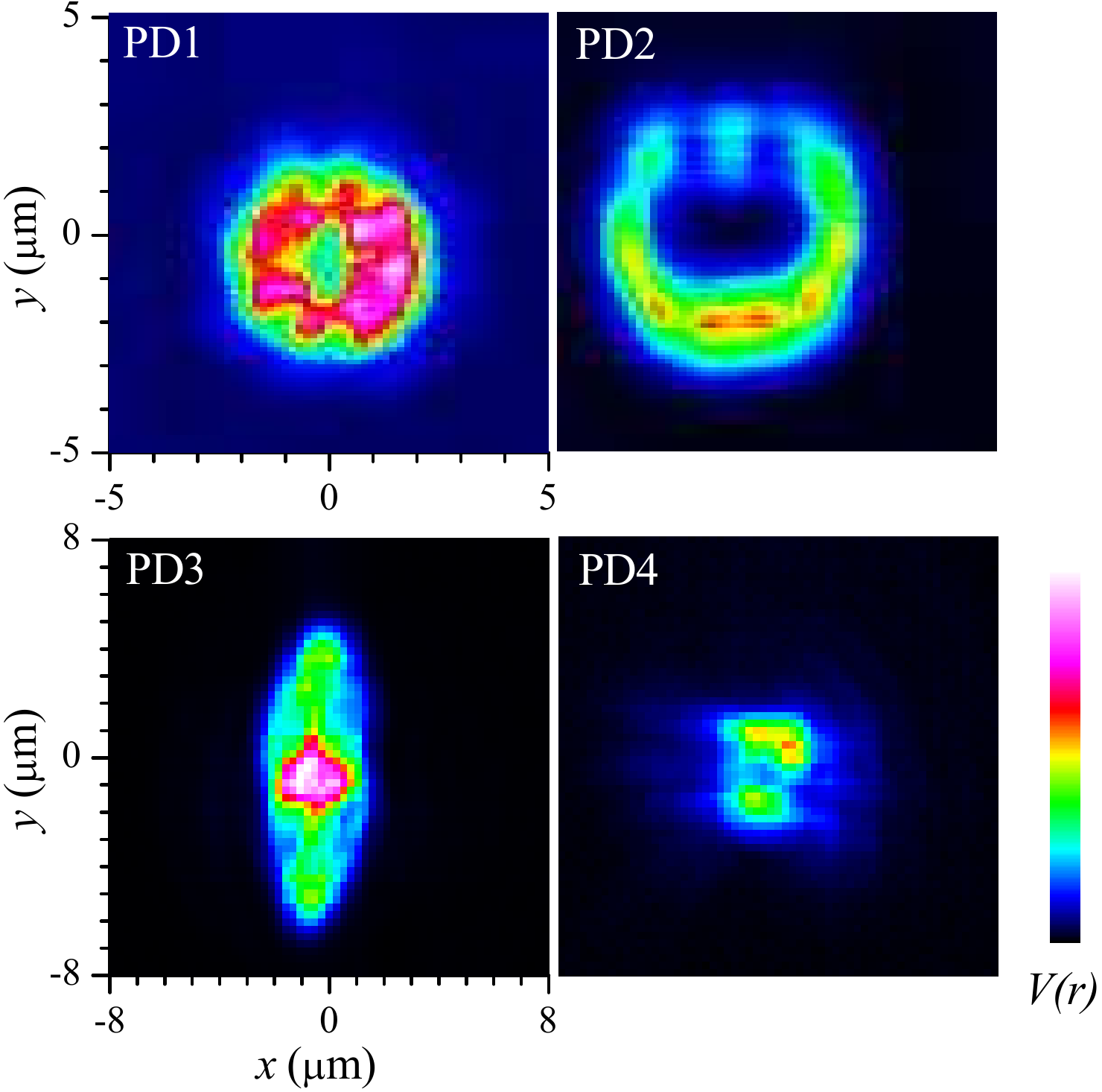}
\caption{Potentials $V_m(\br)$ of PDs calculated using \Eq{eq:potentials}. The color scale is
given, covering 0 to -24\,meV for PD1 and PD2 and 0 to -46\,meV for PD3 and PD4.}
\label{fig:potentials}
\end{figure}

The structural characterization of PD2 by DIC and FIB/SEM is given in \Fig{fig:D111015012fib}.
The surface profile is similar to PD1, with a size of about $7\,\mu$m diameter. The defect
source is much deeper in the structure than for PD1, in the 23rd period of the DBR below the
cavity, and has a larger height of about $90\,$nm. This height is about twice the nominal
height of the GaAs $\lc/4$ layer, and leads to discontinuities of the DBRs for about 4 layers
above the defect. The deep center depression of the defect is consistent with the ring-shaped
polariton confinement potential shown in \Fig{fig:potentials}.

\subsection{Sample MC2} \label{sec:MC2}
\begin{figure}
\includegraphics[width=\columnwidth]{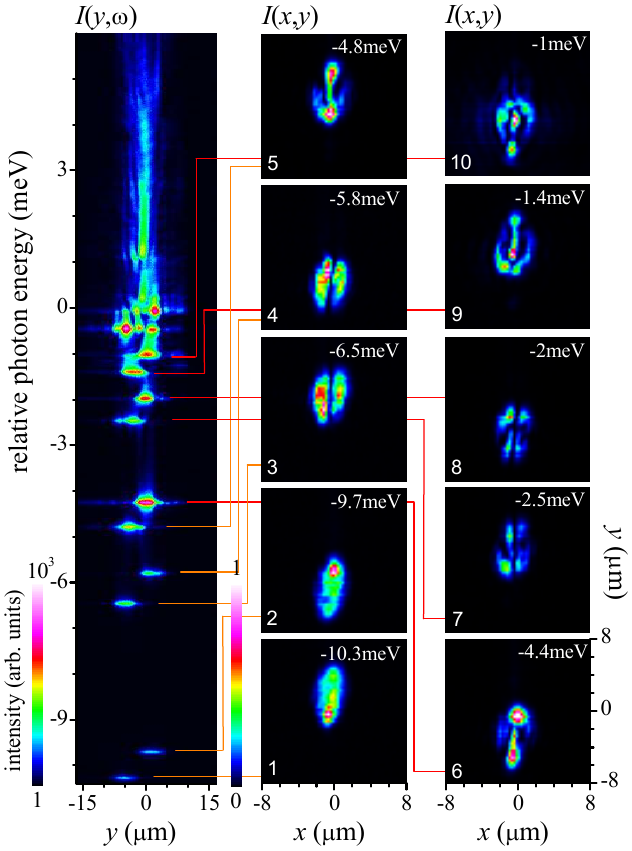}
\caption{Polariton spectrum of PD\,3 with $\hbar\wc=1.4353$\,eV, detailed description as for
\Fig{fig:optical111015018}.} \label{fig:optical110723006}
\end{figure}
\begin{figure}
\includegraphics[width=\columnwidth]{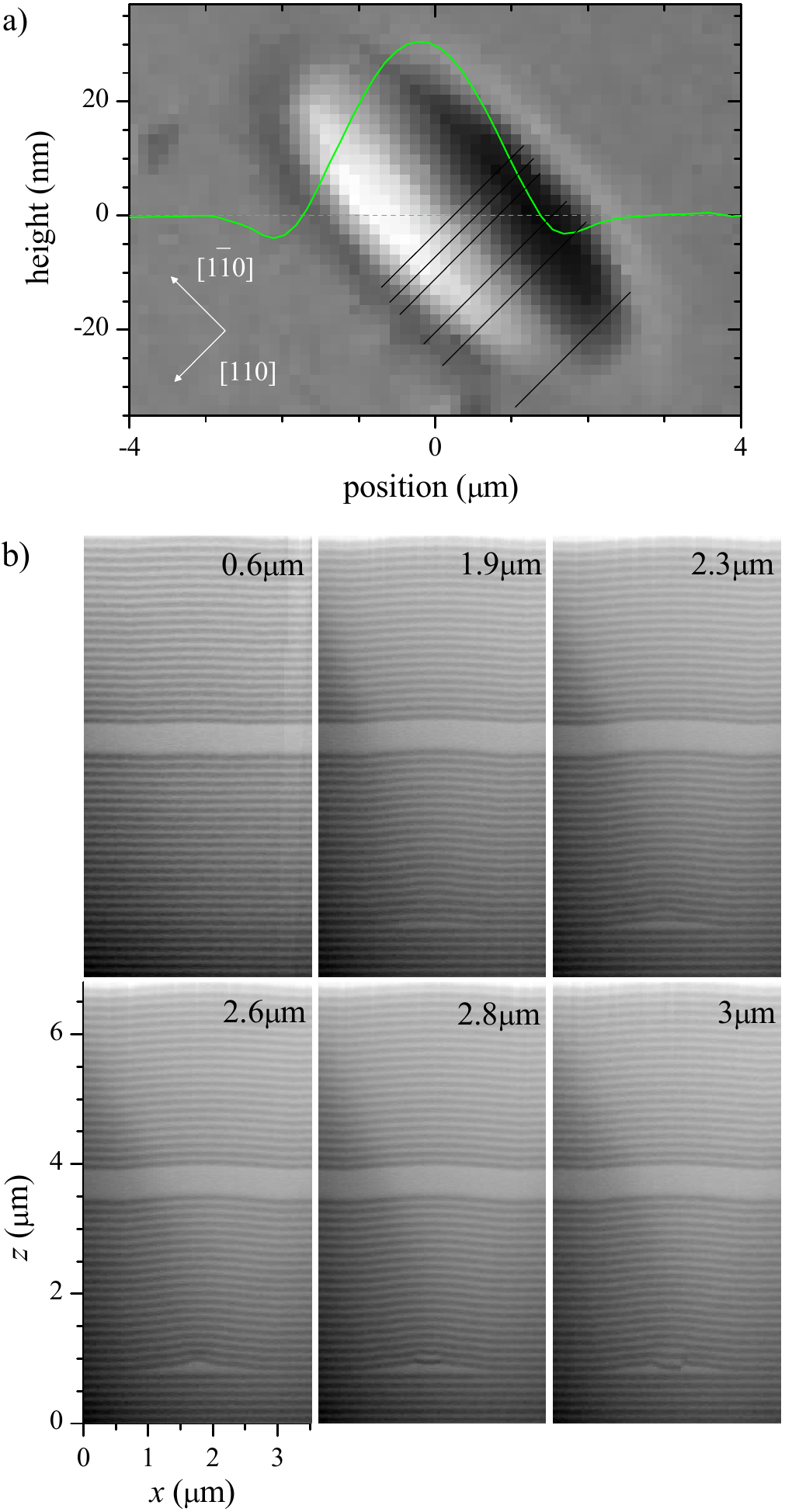}
\caption{Structural characterization of PD\,3, detailed description as for as
\Fig{fig:D111015018fib}.}
%26 top and 21 bottom DBRs is shown.}
\label{fig:D110906006fib}
\end{figure}

This sample was grown at lower temperature than MC1 (see \Tab{tab:Samples}), and shows
oval-shaped defects. Two examples of defects for this samples are given here, referred to as
PD3 and PD4. The polariton states of PD3 are visible in the hyperspectral transmission images
in \Fig{fig:optical110723006}. The states come in nearly degenerate pairs with point reflected
wavefunctions, extended along the x-axis ($[1\bar{1}0]$), for example the state pairs (1,2)
and (3,4), (5,6). The states also have an approximate mirror symmetry about the x axis.
Similar "double-well" eigenstates were observed for several other oval defects in this sample.
All of them exhibit the same sequence of states, while the number of confined states was
varying.

Localization of the states close to the center of the defect, and the absence of mixed states
of different parities indicate a very high potential barrier in the middle, and deep wells on
both sides, with the corresponding potential in $y$ direction could be written as $V(y)\propto
\delta(y/b)+ a/(|y/b|+1)$  with the scaling constants $a,b$. The resulting states of the two
sides $y \lessgtr 0$ do not mix significantly. The states for $y<0$ are shifted by $<1$\,meV
to higher energies. In both wells we observe a ground state, followed by the first excited
state some $4$\,meV above having one node and the second excited state some $2$\,meV further
having two nodes, the third excited state some $2$\,meV with 3 nodes, and higher states.

The structural characterization of PD3 is given in \Fig{fig:D110906006fib}.  The surface
profile is oval, with the extension along $[110]$ of $3\,\mu$m, reduced by a factor of two
compared to the one of PD1, while the extension along $[1\bar{1}0]$ of $6\,\mu$m is similar to
one of PD1. The height of the surface modulation is about 140\,nm, twice the value seen for
PD1 and PD2, and  about twice the nominal height of the GaAs Bragg layer. The height increase
is consistent with the reduced lateral size when accommodating the same volume. The defect
source is in the 20th DBR period below the cavity.
\begin{figure*}
\includegraphics[width=0.95\textwidth]{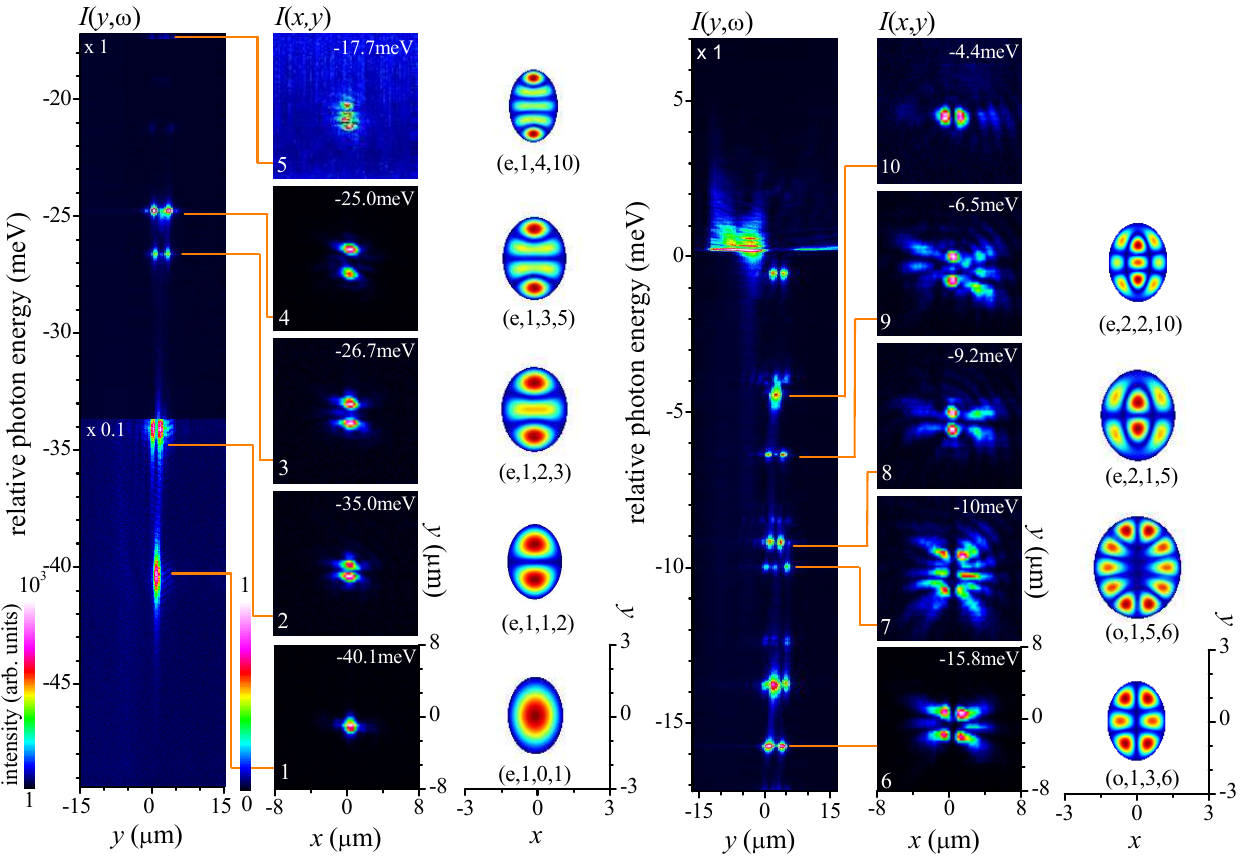}
\caption{Polariton spectrum of PD\,4 with $\hbar\wc=1.4343$\,eV, detailed description as for
\Fig{fig:optical111015018}.} \label{fig:optical110723001}
\end{figure*}
%\
\begin{figure}
\includegraphics[width=\columnwidth]{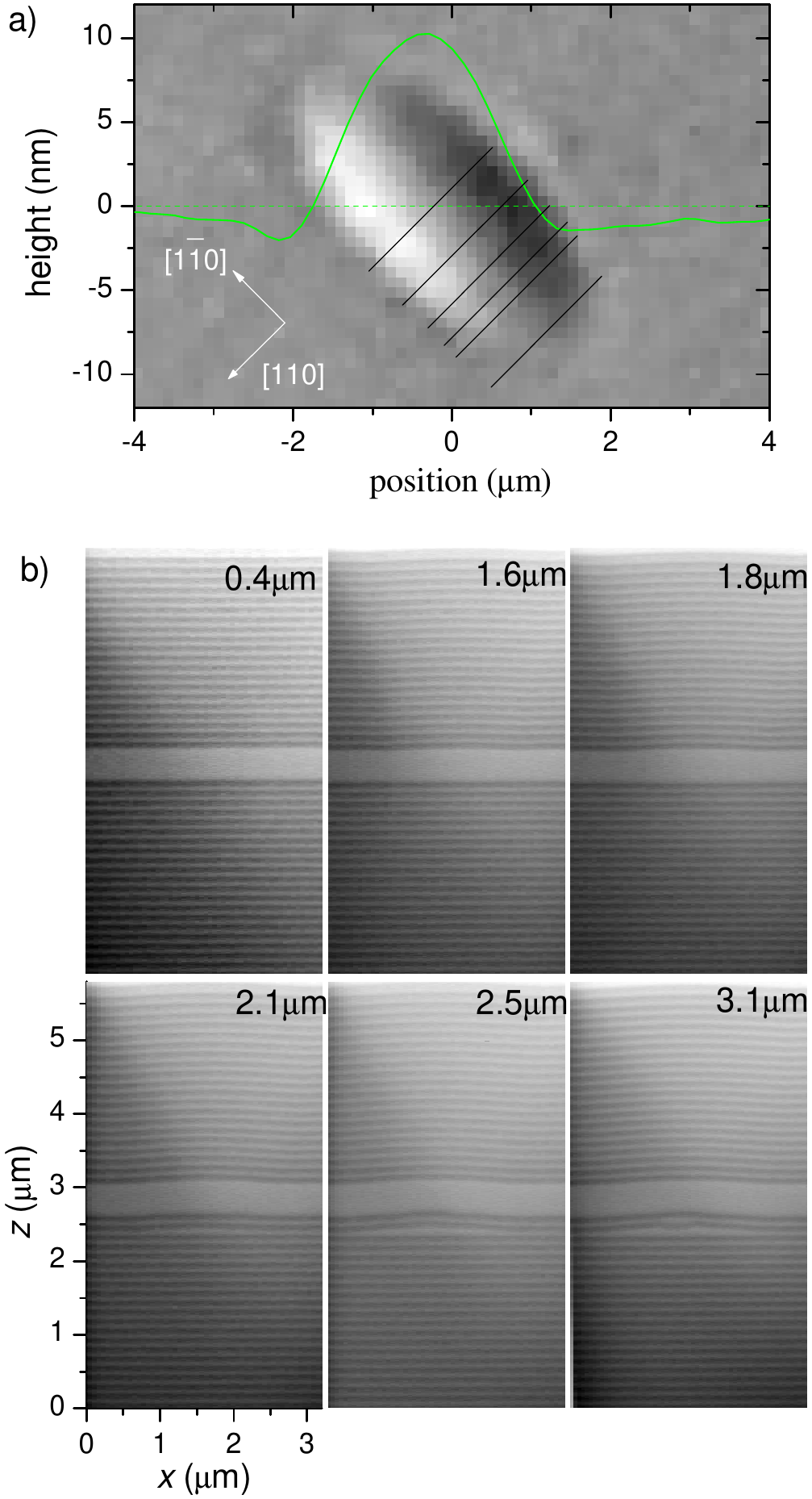}
\caption{Structural characterization of PD\,4, detailed description as for as
\Fig{fig:D111015018fib}.}
%21 top and bottom DBRs is shown.
\label{fig:D110906001fib}
\end{figure}
%\

We now move to the second defect PD4 on MC2, for which the hyperspectral transmission images are
given in \Fig{fig:optical110723001}. It shows the deepest localized states of all PDs studied, with
the ground state $40\,$meV below the continuum. The states show an approximate mirror symmetry
along the $x$ and $y$ axis.
%\
We can model them with Mathieu functions as shown in \Fig{fig:optical110723001}, using a strong
ellipticity. The structural characterization of PD4 is given in \Fig{fig:D110906001fib}. The
surface profile is oval as PD3, but with a 20\% smaller extension and a three time smaller height.
The defect source is in second period of the DBR below the cavity, and has a height of about
$100\,$nm. Being so close to the cavity, the additional GaAs has a strong influence on the
polariton states, and the crater in the middle gives rise to a confinement potential with a barrier
between the center and the peripheral area, as evidenced by the spatial distribution of the
confined wavefunctions.
\subsection{Potential Reconstruction}
\label{sec:potential}
The observed localized polariton states can be related to an effective confinement potential
$V_m(\br)$  for the in-plane polariton motion. We can estimate $V_m(\br)$ using the spectrally
integrated density of states $D_m(\br)$ created by $V_m(\br)$ below the continuum edge as
introduced in \Onlinecite{ZajacPRB12}. We use
\be D_m(\br)=\sum_{n=1}^{n_{\rm m}} |\Psi_n(\br)|^2 \ee
where the bound state probability densities $|\Psi_n(\br)|^2$ are taken as the normalized measured
intensities
\be |\Psi_n(\br)|^2=\frac{I(\br,\omega_n)}{\int I(\br,\omega_n) d\br^{2}}. \label{eq:potentials} \ee
This expression assumes that the emitted field is proportional to the polariton wavefunction,
which is valid for a cavity lifetime which is constant for the in-plane wavevector components
of the bound states. This is a good approximation for small in-plane wavevectors, less than
10\% of the light wavevector  in the cavity of about $26/\mu$m. Some of the strongly localized
states in our study with small feature sizes are likely to deviate from this approximation.
$D_m(\br)$ is given by the integral of the free density of states from zero kinetic energy at
the potential floor to the continuum when neglecting the spatial variation of the confinement
potential, {\em i.e.} in the limit of small level splitting compared to the confinement
potential. In two dimensions the density of states is constant and given by $D_{\rm
2D}=m/(2\pi\hbar^2)$,  and we find $V_m(\br)=-D(\br)/D_{\rm 2D}$. We use the effective mass of
the polaritons from the measured dispersion $m=2\cdot10^{-5}\,m_{\rm e}$, where $m_{\rm e}$ is
the free electron mass. The resulting confinement potentials for the investigated PDs are
shown in Fig.\,\ref{fig:potentials}. The symmetry of the potentials reflect the symmetry of
the localized states.
\subsection{Surface Reconstruction}
\label{sec:SurfRecc}
\begin{figure}
\includegraphics[width=\columnwidth]{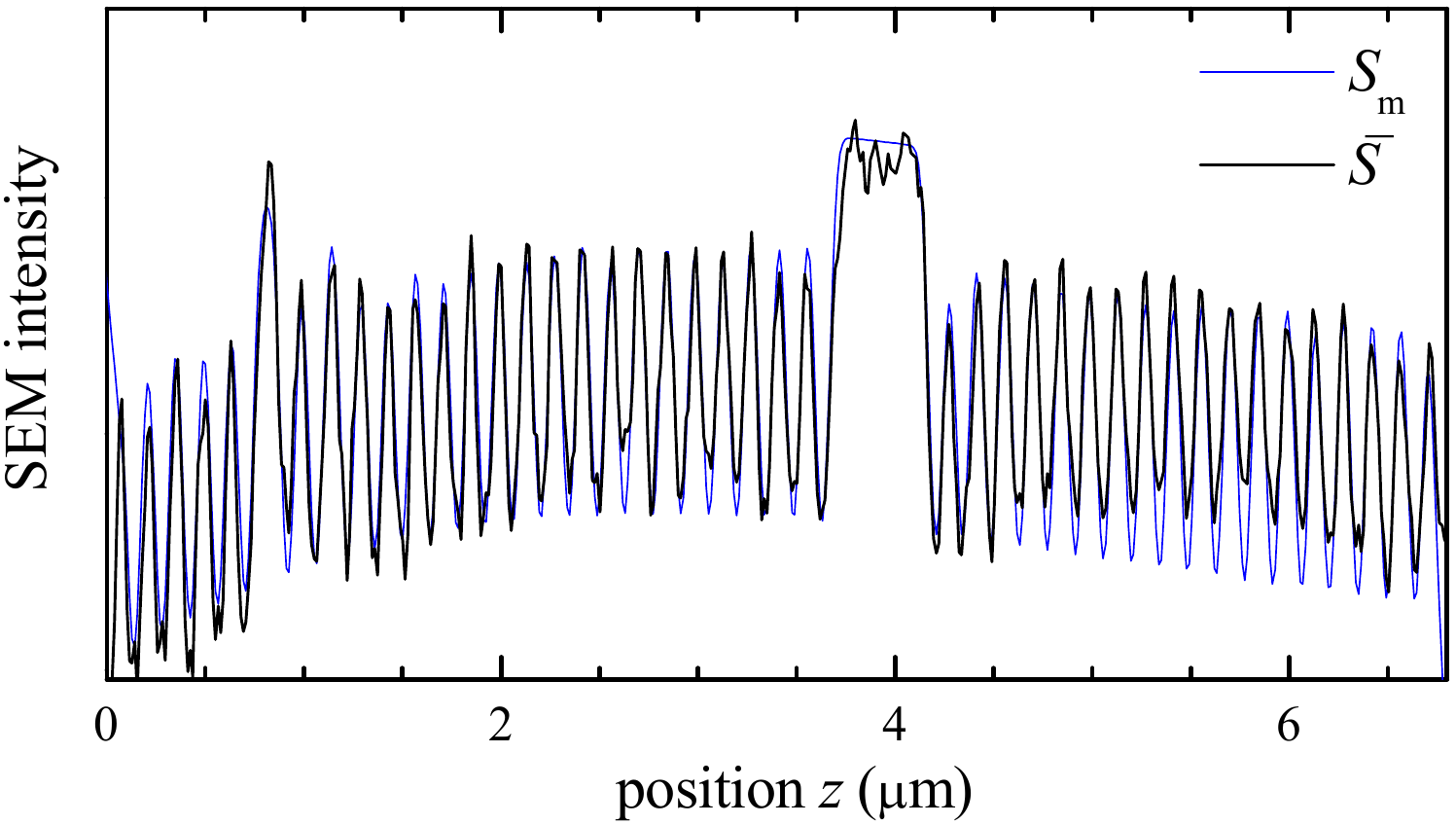}
\caption{Example of a fit (blue line) $\Sf(z)$ to the SEM profile $\bar{S}(z)$ (black line) of
PD3, position $(x,y_n)=(1.5\,\mu$m$,2.5\,\mu$m$)$ in \Fig{fig:D110906006fib}. A linear offset
has been subtracted for better visibility.} \label{fig:2Dfit}
\end{figure}

The series of SEM cross-section images $S(x,y_n,z)$ taken at various $y_n$ (see Figs.
\ref{fig:D111015018fib}, \ref{fig:D111015012fib}, \ref{fig:D110906006fib},
\ref{fig:D110906001fib}), provide volume information of the defects. To reconstruct the shape
of the defects in three dimensions, we determine the position of the interfaces between the
GaAs and AlAs layers in the SEM image. We use PD3 (see Fig.\ref{fig:D110906006fib}) as an
example here. The SEM images show a signal $S$, which is proportional to the detected
secondary electron current, differing by about 10\% between AlAs and GaAs surfaces. The RMS
noise in $S$ was about 5\% of the GaAs signal. SEM images were taken with a nominal
magnification between 65600 and 79500. We calibrated the vertical ($y$) axis to match the
nominal Bragg period, yielding pixel sizes between $10$\,nm and $12$\,nm with 2\% error.
%px error (%) = 10px, for more details see
%origin file with analyze in R:\EPSRC_Superfluidity\Publications\PRB_Oval_defects\Latest\fig
%
The noise of $S$ is limiting the precision with which the layer interface positions can be
determined. To enable a reliable fit of the interface positions, we have averaged the data
over 5 pixels (60\,nm) along $x$, orthogonal to the growth direction $z$. The resulting
$\bar{S}(x,y_n,z)$ was fitted with a model function $\Sf(z)$ of the epitaxial structure along
$z$.
The model assumes a Gaussian resolution of the imaging with a variance of $r/\sqrt{2}$, and a
constant thickness $d$ of the $\lc/4$ AlAs layers, not affected by the defect, which is
motivated by the small surface diffusion length of AlAs compared to GaAs. The sequence of $L$
AlAs layers in GaAs is then described by
\bea
\Sf(z) & = &  a_0 + a_1z + a_2z^2+ A \times  \\
& & \sum_{l=1}^{L}\left[ \mbox{erf} \left( \frac{z - z_{l}}{r} \right) -\mbox{erf} \left (\frac{z -
z_{l} - d}{r} \right) \right] \label{eq:SEMfit}. \nonumber \eea
The polynomial coefficients $a_{0,1,2}$ describe the background, $A$ is half the signal
difference between AlAs and GaAs, and $z_l$ are the positions of the lower interfaces of the
AlAs layers. The layer index $l$ is the AlAs layer number in growth direction. The topmost 3-4
layers were excluded from the fitted region as the background varied strongly due to the
change in secondary electron collection efficiency (see {\em e.g.} \Fig{fig:D111015018fib}b).
The resolution parameter $r$ was 40\,nm corresponding to a FWHM of 67\,nm.
%FWHM = \sigma 2 sqrt(2 ln(2)) = r*2*sqrt(ln2)
The fitted layer positions $z_l$ show a noise of a few nm. An example of such a fit is given in
\Fig{fig:2Dfit}.
Using the $z_l(x,y_n)$ for the different cross-sections $n$, we can reconstruct height maps of
the AlAs layers within the structure across $x$ and $y$. A linear slope and an offset along
$x$ were subtracted from each $z_l(x,y_n)$ of an individual cross-section $n$ to reproduce the
nominal position outside the defect.
\begin{figure}
\includegraphics[width=\columnwidth]{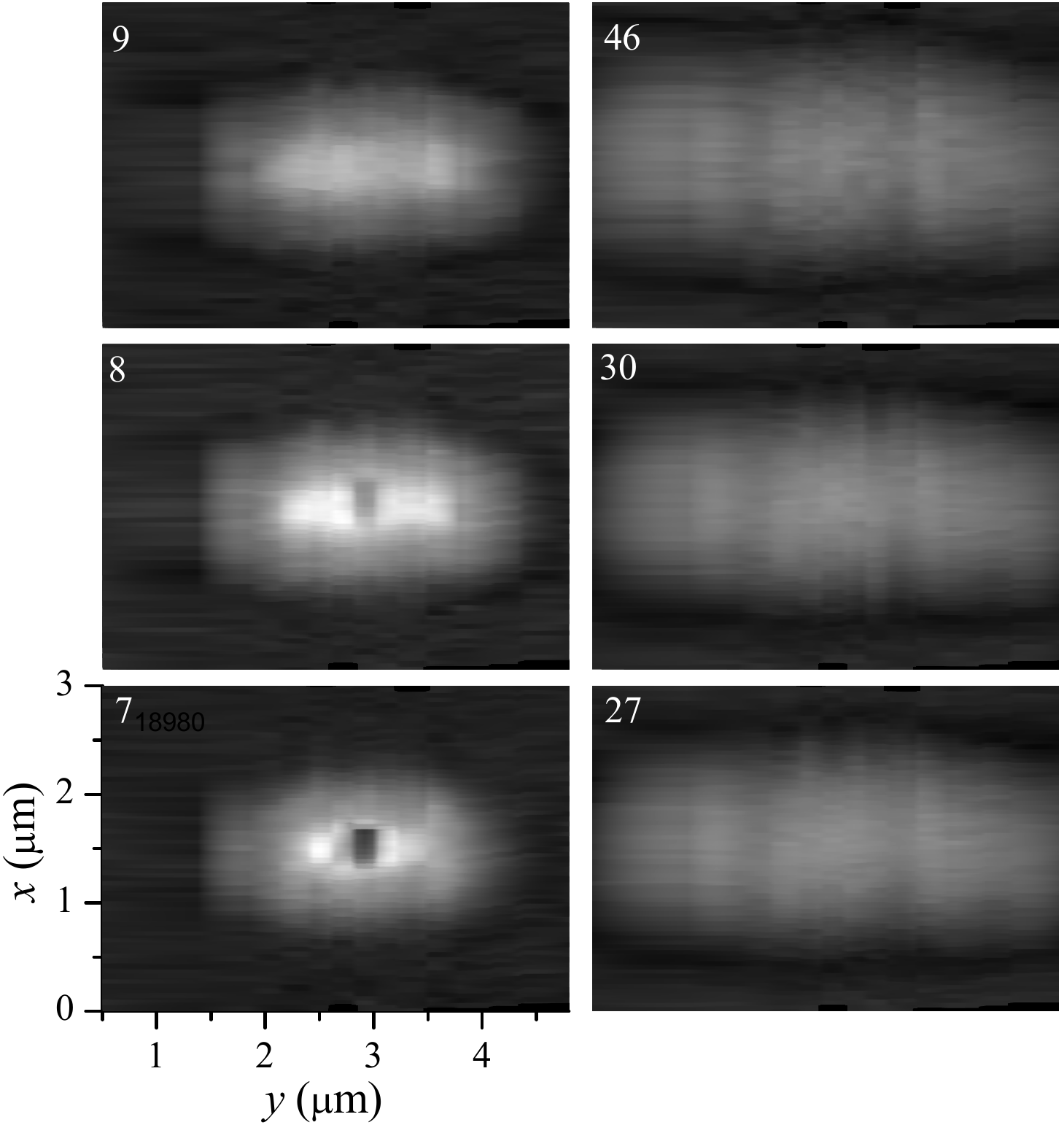} %use 3D_recV2.eps for plot without V
\caption{Height maps of AlAs layers in PD3. The sequential numbers of the layers $n$ are
given. Linear grey scale from -20\,nm (black) and +140\,nm (white) relative to the nominal
layer position. The layer 7 is the first above the GaAs layer containing the droplet, layer 27
is the first layer above the $2\lambda$ GaAs cavity layer, and layer 46 is the last fitted
layer.} \label{fig:3D_rec}
\end{figure}

The height maps of PD3 reconstructed from 15 cross-sections (see lines in
\Fig{fig:D110906006fib}, not all shown) are displayed in \Fig{fig:3D_rec}. The evolution of
the surface modulation can be followed. The first layer above the defect source ($l=7$) shows
the center depression of the GaAs, similar to what observed in liquid droplet epitaxy
\cite{ManoNL05}. Two maxima of the thickness are observed along the preferential surface
diffusion direction $[1\bar{1}0]$. With increasing layer number, first the depression
disappears ($l=9$), followed by a general extension and flattening of the structure. By
integrating height profiles of defect for different cuts we have determined the volume of the
additional GaAs material as constant within the error. From this volume, we can deduce the
radius of the deposited Ga droplet which is $86\pm6$\,nm.

\section{Discussion}
\label{sec:disc} For all of the 15 PDs investigated in this work, of which four have been
shown as examples, we find a similar origin - a local increase in a GaAs layer thickness with
a depression in the center. The additional GaAs volume can only be created by a local
deposition of Ga, as the growth is limited by the group III element, while the group V element
As is provided in a much larger amount given by the V/III flux ratio of about 50, and desorbs
if not bound to the surface with a Ga atom to form GaAs. The only available source for this
excess Ga deposition are Ga droplets from the Ga source.

The shapes of the polariton potentials created by the PD are a consequence of the Ga droplet
size, and its deposition position relative to the cavity layer. In order to simulate the
0-dimensional polariton states quantitatively, a full three-dimensional simulation of the mode
structure in the cavity would be needed, which is beyond the scope of the present work. For a
qualitative argument, one can use a first-order perturbation picture. The polariton intensity
is decaying exponentially into the Bragg mirror with a decay length of about 400\,nm. For PD2,
the GaAs layer thickening is $50\,$nm, but it is separated by 23 DBR periods, about $3\,\mu$m,
or 8 decay lengths from the cavity, where the polariton intensity has decayed to 0.02\%. This
results in a small influence to the polaritons and a small localization energy of the ground
state of 5\,meV. For this defect we observe a large central crater and DBR layer
discontinuities, as can be seen on the central cut in \Fig{fig:D111015018fib}. This could give
rise to a repulsive central part of the PD2 potential as observed in \Fig{fig:potentials}.

In PD1, the Ga droplet had a similar size as in PD2 (GaAs thickening 90\,nm), but it hit the
surface only three DBR periods above the cavity layer. Even though the induced structural
perturbation propagates away from the cavity layer, is has a much larger influence, with the
third excited state at -9\,meV and an estimated ground state confinement energy of 20\,meV.
The significantly smaller lateral extension of the defects in MC2 leads to a larger height of
the perturbation, which in turn results in stronger confinement of polariton states with
shapes as seen for PD3 and PD4.

The evolution of the defect structure during growth can be pictured as follows.  After a Ga
droplet was deposited on the surface, Ga diffuses over the surface from the droplet to the
surrounding areas. Due the large V-III flux ratio, there is sufficient surplus of As$_2$
impinging onto the surface to convert the diffusing Ga into GaAs, leading to an additional
GaAs thickness which decays with the distance from the deposition spot, according to the Ga
diffusion length. The depression in the center of the resulting profile is due to reduced GaAs
growth below the Ga droplet, which requires the diffusion of As through the Ga droplet to the
GaAs surface. Once the droplet has been consumed, the subsequent GaAs growth generally tends
to smooth the surface due to the Ga surface diffusion and the preferential attachment of Ga at
monolayer steps, which have a density proportional to the surface gradient for gradients
superseding the gradient due to monolayer islands (for a island size of 20\,nm a gradient of
1\%). Al instead has a much shorter diffusion length, and therefore the surface profile is
essentially conserved during the growth of the AlAs layers. For Ga grown on (001) oriented
substrate at $590^{\circ}$C at a V/III flux ratio of 2 and a growth rate of $0.25\,\mu$m/h the
diffusion length was reported\cite{KoshibaAPL94} to be $1\,\mu$m and $0.02\,\mu$m for Ga and
Al, respectively.

The observed PD anisotropy of 1:2 to 1:3 along the $[110]$:$[1\bar{1}0]$ directions for MC2
grown at a temperature of $590^\circ$C reduces to less than 1:1.1 for MC1 grown at
$715^\circ$C. This finding can be explained by temperature dependent diffusion lengths $D$ for
these two crystallographic directions. In \Onlinecite{OhtaJCG89} $D_{[1\bar{1}0]} =
4D_{[110]}$ was found resulting in diffusion lengths of $L_{[1\bar{1}0]} = 2L_{[110]}$ for a
V/III flux ratio of 1.5 and growth temperatures in the range of $600^{\circ}$C, in agreement
with the aspect ratio of the defects found in MC2. The reduction of the anisotropy for the
higher growth temperature of MC1 indicates an activated diffusion with different activation
energies in the two directions. At higher temperatures, the thermal energy supersedes the
activation energies and a kinetically limited isotropic diffusion is recovered. The presence
of different activation energies for diffusion in the two crystallographic directions is
plausible as during growth the GaAs surface shows a reconstruction\cite{BiegelsenPRB90} giving
rise to a channel-like structure along $[1\bar{1}0]$.

\section{Conclusions}
\label{sec:sum} We have shown that oval or round defects in MBE grown GaAs microcavities
create zero-dimensional polariton states of narrow linewidths. We have revealed their
three-dimensional structure and their formation mechanism, an impinging Ga droplet during
growth. While we have deduced effective confinement potentials for the defects, a quantitative
modeling of the polariton spectra from the three-dimensional structural information obtained
by the FIB/SEM data is presently missing. In the context of polaritonic
devices,\cite{LiewPhysicaE11} our work indicates an approach to manufacture two-dimensional
polaritonic traps by intentional creation of Ga droplets at a specific position during the MBE
growth of a microcavity, rather than ex-situ etching as described in \Onlinecite{CernaPRB09}.
One could also use Ga droplet epitaxy \cite{MantovaniJAP04} with a low density to create
well-defined localized polariton states in microcavities. The narrow linewidths of the
polariton states formed in this way are favorable for zero-dimensional polariton
switches.\cite{ParaisoNatMatt10}

\section{Acknowledgments}
The samples were grown at the EPSRC National Centre for III-V Technologies, Sheffield, UK, by
Maxime Hugues and Mark Hopkinson (MC1), and Edmund Clarke (MC2). The FIB/SEM investigations
were conducted at the London Centre for Nanotechnology, and funded by the EPSRC Access to
Materials Research Equipment Initiative under grant EP/F019564/1. We thank Suguo Huo and Paul
Warburton for training and assistance with the FIB/SEM. We acknowledge discussions with Paola
Atkinson and Edmund Clarke on the growth kinetics, and help with the photolithography by Phil
Buckle and Karen Barnett. This work was supported by the EPSRC under grant n. EP/F027958/1.

\section{Appendix}
\subsection{Quantitative Differential Interference Contrast Microscopy}

\newcommand{\Shear}{\mathbf{\Delta}}
\newcommand{\phioff}{\phi_{\rm o}}
\newcommand{\IDIC}{I_{\rm DIC}}
\newcommand{\CDIC}{C_{\rm DIC}}

Differential interference contrast microscopy (DIC), also know as Nomarski microscopy, was used in
reflection in this experiment. A Nomarski prism assembly (DIC Slider U-DICT with Polarizer U-ANT)
is mounted in a Olympus BX-50 upright microscope. The illumination from a mercury arc lamp is split
by the Normarki prism into two beams 1,2 shifted by the shear displacement $\Shear$ in the object
plane, with linear polarizations along and orthogonal to $\Shear$. The reflected beams are
recombined by the prism, creating a polarization state depending on their relative phase
$\varphi_1-\varphi_2$. The transmission through the polarizer depends on the polarization state,
such that the intensity depends on the relative phase in the way
\be 2\IDIC=I\left(1-\cos\left(\phioff+\varphi_1-\varphi_2\right)\right) \label{eq:DICI}\ee
where the offset phase $\phioff$ is introduced by an adjustable spatial offset of the
Normarski prism along the optical axis from its nominal position for which the beams are not
displaced in the directional space (objective back focal plane). The shear $\Shear$ is similar
to the optical resolution of the microscope objective, which allows to approximate the phase
difference between the two beams in first order as the shear times the phase gradient at the
observed position, $\varphi_1-\varphi_2\approx\Shear\cdot\nabla\varphi$, such that
\Eq{eq:DICI} can be written as
$2\IDIC=I\left(1-\cos\left(\phioff+\Shear\cdot\nabla\varphi\right)\right)$. Choosing
$\phioff=\pm\pi/2$, and developing up to first order in the phase difference, we get
$2\IDIC^\pm=I\left(1\pm\Shear\cdot\nabla\varphi\right)$. Measuring $\IDIC$ for both offset
phases, we determine the contrast
\begin{figure}
\vspace{0.5cm}
\includegraphics[width=\columnwidth]{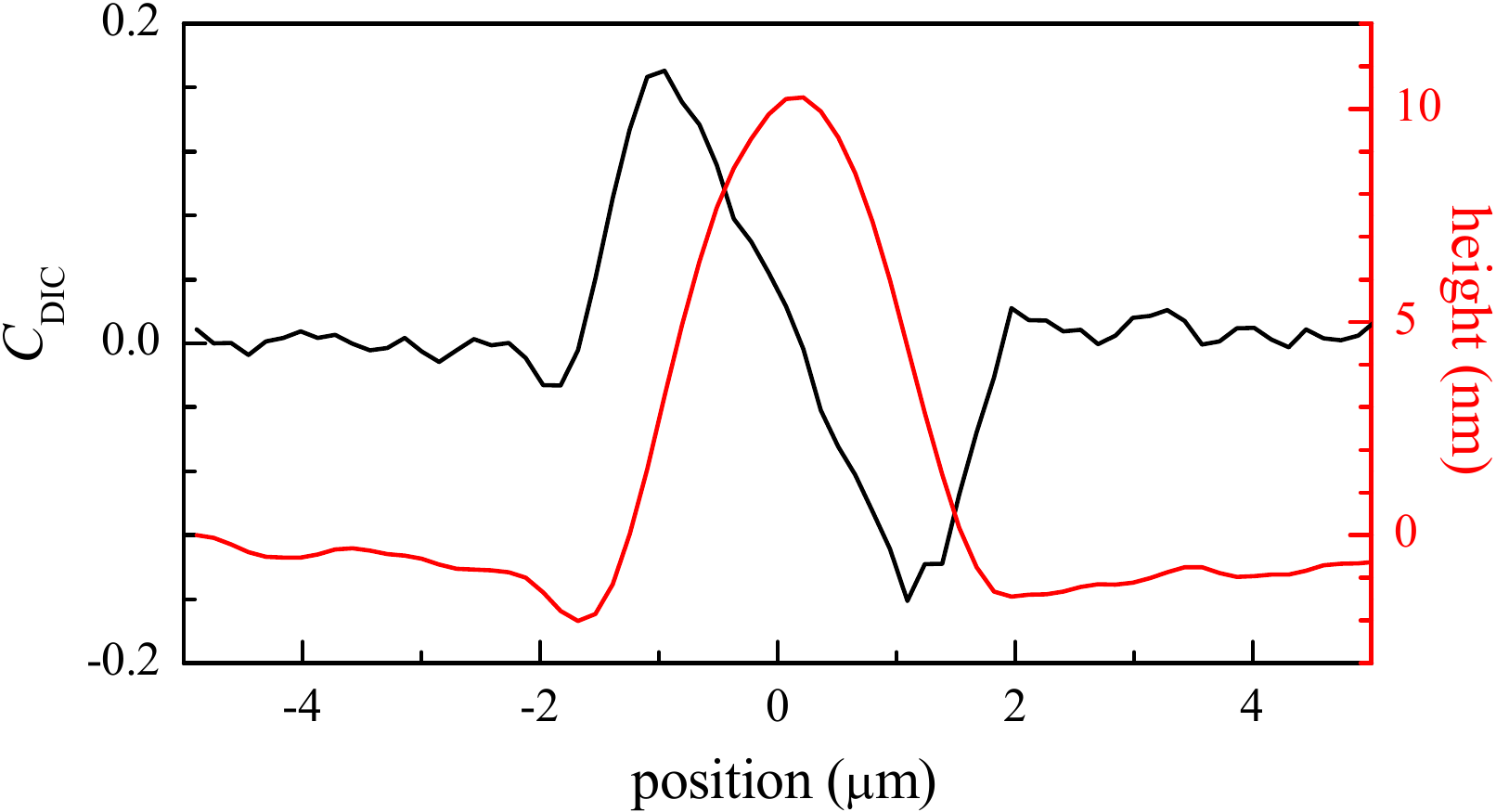}
\caption{DIC contrast $\CDIC$ (black line) as function of the sample position $s$ along the shear
direction, and resulting height profile $h$ (red line) calculated using \Eq{eq:DICh}.}
\label{fig:DIC_principle}
\end{figure}
\be \CDIC=\frac{\IDIC^+ - \IDIC^-}{\IDIC^+ + \IDIC^-}=\Shear\cdot\nabla\varphi \ee
We can now integrate the contrast along the shear direction to retrieve the phase $\varphi$.
In reflection, the phase is related to the surface height $h$ by $\varphi=4\pi h/\lambda$ with
the wavelength $\lambda$ of the light, such that we arrive at
\be h(s) = \frac{\lambda}{4\pi|\Shear|}\int_0^s\CDIC ds' \label{eq:DICh}\ee
We assumed here that sample is not birefringent and that the phase shift of the reflected
light is given by the height of the sample surface only, neglecting internal interfaces. The
latter is justified as the green light is absorbed strongly by the structure. The height
$h(s)$ was determined using \Eq{eq:DICh} with $s$ along the direction of the shear $\Shear$.
In the measurements presented in this work we used a UplanFL 20x/0.5NA objective, for which
the shear was determined to be $|\Shear|=0.55\,\mu$m using a calibration slide in transmission
DIC consisting of a PMMA pattern of a 200\,nm thickness on a glass coverslip, in which case
the phase is given by $\varphi=2\pi h(n_1-n_2)$ with the refractive index difference
$n_1-n_2=0.48$ between PMMA and air. To compensate for systematic errors, the measured $\CDIC$
across the center of the defect was corrected by the $\CDIC$ along a line displaced
perpendicular to the shear, just outside of the defect. An examples of a measured $\CDIC$ and
the resulting height profile $h(s)$ for defect PD4 are shown in \Fig{fig:DIC_principle}.
%
%\bibstyle{prbsty}
%\bibliography{JZthesis,langsrv}

\begin{thebibliography}{32}
\expandafter\ifx\csname natexlab\endcsname\relax\def\natexlab#1{#1}\fi
\expandafter\ifx\csname bibnamefont\endcsname\relax
  \def\bibnamefont#1{#1}\fi
\expandafter\ifx\csname bibfnamefont\endcsname\relax
  \def\bibfnamefont#1{#1}\fi
\expandafter\ifx\csname citenamefont\endcsname\relax
  \def\citenamefont#1{#1}\fi
\expandafter\ifx\csname url\endcsname\relax
  \def\url#1{\texttt{#1}}\fi
\expandafter\ifx\csname urlprefix\endcsname\relax\def\urlprefix{URL }\fi
\providecommand{\bibinfo}[2]{#2}
\providecommand{\eprint}[2][]{\url{#2}}

\bibitem[{\citenamefont{Kasprzak et~al.}(2006)\citenamefont{Kasprzak, Richard,
  Kundermann, Baas, Jeambrun, Keeling, Marchetti, Szymanska, Andre, Staehli
  et~al.}}]{KasprzakNature06}
\bibinfo{author}{\bibfnamefont{J.}~\bibnamefont{Kasprzak}},
  \bibinfo{author}{\bibfnamefont{M.}~\bibnamefont{Richard}},
  \bibinfo{author}{\bibfnamefont{S.}~\bibnamefont{Kundermann}},
  \bibinfo{author}{\bibfnamefont{A.}~\bibnamefont{Baas}},
  \bibinfo{author}{\bibfnamefont{P.}~\bibnamefont{Jeambrun}},
  \bibinfo{author}{\bibfnamefont{J.~M.~J.} \bibnamefont{Keeling}},
  \bibinfo{author}{\bibfnamefont{F.~M.} \bibnamefont{Marchetti}},
  \bibinfo{author}{\bibfnamefont{M.~H.} \bibnamefont{Szymanska}},
  \bibinfo{author}{\bibfnamefont{R.}~\bibnamefont{Andre}},
  \bibinfo{author}{\bibfnamefont{J.~L.} \bibnamefont{Staehli}},
  \bibnamefont{et~al.}, \bibinfo{journal}{Nature}
  \textbf{\bibinfo{volume}{443,}}, \bibinfo{pages}{409} (\bibinfo{year}{2006}).

\bibitem[{\citenamefont{Lagoudakis et~al.}(2008)\citenamefont{Lagoudakis,
  Wouters, Richard, Baas, Carusotto, André, Dang, and
  Deveaud-Pledran}}]{LagoudakisNatPhys08}
\bibinfo{author}{\bibfnamefont{K.~G.} \bibnamefont{Lagoudakis}},
  \bibinfo{author}{\bibfnamefont{M.}~\bibnamefont{Wouters}},
  \bibinfo{author}{\bibfnamefont{M.}~\bibnamefont{Richard}},
  \bibinfo{author}{\bibfnamefont{A.}~\bibnamefont{Baas}},
  \bibinfo{author}{\bibfnamefont{I.}~\bibnamefont{Carusotto}},
  \bibinfo{author}{\bibfnamefont{R.}~\bibnamefont{André}},
  \bibinfo{author}{\bibfnamefont{D.~L.~S.} \bibnamefont{Dang}},
  \bibnamefont{and}
  \bibinfo{author}{\bibfnamefont{B.}~\bibnamefont{Deveaud-Pledran}},
  \bibinfo{journal}{Nature Physics} \textbf{\bibinfo{volume}{4}},
  \bibinfo{pages}{706} (\bibinfo{year}{2008}).

\bibitem[{\citenamefont{Amo et~al.}(2009)\citenamefont{Amo, Sanvitto, Laussy,
  Ballarini, del Valle, Martin, Lemaître, Bloch, Krizhanovskii, Skolnick
  et~al.}}]{AmoNature09}
\bibinfo{author}{\bibfnamefont{A.}~\bibnamefont{Amo}},
  \bibinfo{author}{\bibfnamefont{D.}~\bibnamefont{Sanvitto}},
  \bibinfo{author}{\bibfnamefont{F.~P.} \bibnamefont{Laussy}},
  \bibinfo{author}{\bibfnamefont{D.}~\bibnamefont{Ballarini}},
  \bibinfo{author}{\bibfnamefont{E.}~\bibnamefont{del Valle}},
  \bibinfo{author}{\bibfnamefont{M.~D.} \bibnamefont{Martin}},
  \bibinfo{author}{\bibfnamefont{A.}~\bibnamefont{Lemaître}},
  \bibinfo{author}{\bibfnamefont{J.}~\bibnamefont{Bloch}},
  \bibinfo{author}{\bibfnamefont{D.~N.} \bibnamefont{Krizhanovskii}},
  \bibinfo{author}{\bibfnamefont{M.~S.} \bibnamefont{Skolnick}},
  \bibnamefont{et~al.}, \bibinfo{journal}{Nature}
  \textbf{\bibinfo{volume}{457}}, \bibinfo{pages}{291} (\bibinfo{year}{2009}).

\bibitem[{\citenamefont{Zajac et~al.}(2012{\natexlab{a}})\citenamefont{Zajac,
  Clarke, and Langbein}}]{ZajacAPL12}
\bibinfo{author}{\bibfnamefont{J.~M.} \bibnamefont{Zajac}},
  \bibinfo{author}{\bibfnamefont{E.}~\bibnamefont{Clarke}}, \bibnamefont{and}
  \bibinfo{author}{\bibfnamefont{W.}~\bibnamefont{Langbein}},
  \bibinfo{journal}{Applied Physics Letters} \textbf{\bibinfo{volume}{101}},
  \bibinfo{eid}{041114} (pages~\bibinfo{numpages}{4})
  (\bibinfo{year}{2012}{\natexlab{a}}).

\bibitem[{\citenamefont{Abbarchi et~al.}(2012)\citenamefont{Abbarchi,
  Diederichs, Largeau, Ardizzone, Mauguin, Lecomte, Lemaitre, Bloch,
  Roussignol, and Tignon}}]{AbbarchiPRB12}
\bibinfo{author}{\bibfnamefont{M.}~\bibnamefont{Abbarchi}},
  \bibinfo{author}{\bibfnamefont{C.}~\bibnamefont{Diederichs}},
  \bibinfo{author}{\bibfnamefont{L.}~\bibnamefont{Largeau}},
  \bibinfo{author}{\bibfnamefont{V.}~\bibnamefont{Ardizzone}},
  \bibinfo{author}{\bibfnamefont{O.}~\bibnamefont{Mauguin}},
  \bibinfo{author}{\bibfnamefont{T.}~\bibnamefont{Lecomte}},
  \bibinfo{author}{\bibfnamefont{A.}~\bibnamefont{Lemaitre}},
  \bibinfo{author}{\bibfnamefont{J.}~\bibnamefont{Bloch}},
  \bibinfo{author}{\bibfnamefont{P.}~\bibnamefont{Roussignol}},
  \bibnamefont{and} \bibinfo{author}{\bibfnamefont{J.}~\bibnamefont{Tignon}},
  \bibinfo{journal}{Phys. Rev. B} \textbf{\bibinfo{volume}{85}},
  \bibinfo{pages}{045316} (\bibinfo{year}{2012}).

\bibitem[{\citenamefont{Zajac et~al.}(2012{\natexlab{b}})\citenamefont{Zajac,
  Langbein, Hugues, and Hopkinson}}]{ZajacPRB12}
\bibinfo{author}{\bibfnamefont{J.~M.} \bibnamefont{Zajac}},
  \bibinfo{author}{\bibfnamefont{W.}~\bibnamefont{Langbein}},
  \bibinfo{author}{\bibfnamefont{M.}~\bibnamefont{Hugues}}, \bibnamefont{and}
  \bibinfo{author}{\bibfnamefont{M.}~\bibnamefont{Hopkinson}},
  \bibinfo{journal}{Phys. Rev. B} \textbf{\bibinfo{volume}{85}},
  \bibinfo{pages}{165309} (\bibinfo{year}{2012}{\natexlab{b}}).

\bibitem[{\citenamefont{Langbein et~al.}(2002)\citenamefont{Langbein, Runge,
  Savona, and Zimmermann}}]{LangbeinPRL02a}
\bibinfo{author}{\bibfnamefont{W.}~\bibnamefont{Langbein}},
  \bibinfo{author}{\bibfnamefont{E.}~\bibnamefont{Runge}},
  \bibinfo{author}{\bibfnamefont{V.}~\bibnamefont{Savona}}, \bibnamefont{and}
  \bibinfo{author}{\bibfnamefont{R.}~\bibnamefont{Zimmermann}},
  \bibinfo{journal}{Phys. Rev. Lett.} \textbf{\bibinfo{volume}{89}},
  \bibinfo{pages}{157401} (\bibinfo{year}{2002}).

\bibitem[{\citenamefont{Gurioli et~al.}(2005)\citenamefont{Gurioli, Bogani,
  Cavigli, Gibbs, Khitrova, and Wiersma}}]{GurioliPRL05}
\bibinfo{author}{\bibfnamefont{M.}~\bibnamefont{Gurioli}},
  \bibinfo{author}{\bibfnamefont{F.}~\bibnamefont{Bogani}},
  \bibinfo{author}{\bibfnamefont{L.}~\bibnamefont{Cavigli}},
  \bibinfo{author}{\bibfnamefont{H.}~\bibnamefont{Gibbs}},
  \bibinfo{author}{\bibfnamefont{G.}~\bibnamefont{Khitrova}}, \bibnamefont{and}
  \bibinfo{author}{\bibfnamefont{D.~S.} \bibnamefont{Wiersma}},
  \bibinfo{journal}{Phys. Rev. Lett.} \textbf{\bibinfo{volume}{94}},
  \bibinfo{pages}{183901} (\bibinfo{year}{2005}).

\bibitem[{\citenamefont{Krizhanovskii et~al.}(2009)\citenamefont{Krizhanovskii,
  Lagoudakis, Wouters, Pietka, Bradley, Guda, Whittaker, Skolnick,
  Deveaud-Plaedran, Richard et~al.}}]{KrizhanovskiiPRB09}
\bibinfo{author}{\bibfnamefont{D.~N.} \bibnamefont{Krizhanovskii}},
  \bibinfo{author}{\bibfnamefont{K.~G.} \bibnamefont{Lagoudakis}},
  \bibinfo{author}{\bibfnamefont{M.}~\bibnamefont{Wouters}},
  \bibinfo{author}{\bibfnamefont{B.}~\bibnamefont{Pietka}},
  \bibinfo{author}{\bibfnamefont{R.~A.} \bibnamefont{Bradley}},
  \bibinfo{author}{\bibfnamefont{K.}~\bibnamefont{Guda}},
  \bibinfo{author}{\bibfnamefont{D.~M.} \bibnamefont{Whittaker}},
  \bibinfo{author}{\bibfnamefont{M.~S.} \bibnamefont{Skolnick}},
  \bibinfo{author}{\bibfnamefont{B.}~\bibnamefont{Deveaud-Plaedran}},
  \bibinfo{author}{\bibfnamefont{M.}~\bibnamefont{Richard}},
  \bibnamefont{et~al.}, \bibinfo{journal}{Phys. Rev. B}
  \textbf{\bibinfo{volume}{80}}, \bibinfo{pages}{045317}
  (\bibinfo{year}{2009}).

\bibitem[{\citenamefont{Chand and Chu}(1990)}]{ChandJCrystGrowth90}
\bibinfo{author}{\bibfnamefont{N.}~\bibnamefont{Chand}} \bibnamefont{and}
  \bibinfo{author}{\bibfnamefont{S.}~\bibnamefont{Chu}},
  \bibinfo{journal}{Journal of Crystal Growth} \textbf{\bibinfo{volume}{104}},
  \bibinfo{pages}{485} (\bibinfo{year}{1990}).

\bibitem[{\citenamefont{Fujiwara et~al.}(1987)\citenamefont{Fujiwara, Kanmoto,
  Ohta, Tokuda, and Nakayama}}]{FujiwaraJCrystGrowth87}
\bibinfo{author}{\bibfnamefont{K.}~\bibnamefont{Fujiwara}},
  \bibinfo{author}{\bibfnamefont{K.}~\bibnamefont{Kanmoto}},
  \bibinfo{author}{\bibfnamefont{Y.}~\bibnamefont{Ohta}},
  \bibinfo{author}{\bibfnamefont{Y.}~\bibnamefont{Tokuda}}, \bibnamefont{and}
  \bibinfo{author}{\bibfnamefont{T.}~\bibnamefont{Nakayama}},
  \bibinfo{journal}{Journal of Crustal Growth} \textbf{\bibinfo{volume}{80}},
  \bibinfo{pages}{104} (\bibinfo{year}{1987}).

\bibitem[{\citenamefont{Orme et~al.}(1994)\citenamefont{Orme, Johnson, Leung,
  and Orr}}]{OrmeMRSSP94}
\bibinfo{author}{\bibfnamefont{C.}~\bibnamefont{Orme}},
  \bibinfo{author}{\bibfnamefont{M.~D.} \bibnamefont{Johnson}},
  \bibinfo{author}{\bibfnamefont{K.~T.} \bibnamefont{Leung}}, \bibnamefont{and}
  \bibinfo{author}{\bibfnamefont{B.~G.} \bibnamefont{Orr}},
  \bibinfo{journal}{Material Research Society Symposium Proceedings Vol 340}
  (\bibinfo{year}{1994}), ISSN \bibinfo{issn}{0921-5107}.

\bibitem[{\citenamefont{Kreutzer et~al.}(1999)\citenamefont{Kreutzer, Zacher,
  Naumann, Franke, and Anton}}]{KreutzerUltramicroscopy99}
\bibinfo{author}{\bibfnamefont{P.}~\bibnamefont{Kreutzer}},
  \bibinfo{author}{\bibfnamefont{T.}~\bibnamefont{Zacher}},
  \bibinfo{author}{\bibfnamefont{W.}~\bibnamefont{Naumann}},
  \bibinfo{author}{\bibfnamefont{T.}~\bibnamefont{Franke}}, \bibnamefont{and}
  \bibinfo{author}{\bibfnamefont{R.}~\bibnamefont{Anton}},
  \bibinfo{journal}{Ultramicroscopy} \textbf{\bibinfo{volume}{76}},
  \bibinfo{pages}{107 } (\bibinfo{year}{1999}), ISSN \bibinfo{issn}{0304-3991}.

\bibitem[{\citenamefont{Kawada et~al.}(1993)\citenamefont{Kawada, Shirayone,
  and Takahashi}}]{KawadaJCG93}
\bibinfo{author}{\bibfnamefont{H.}~\bibnamefont{Kawada}},
  \bibinfo{author}{\bibfnamefont{S.}~\bibnamefont{Shirayone}},
  \bibnamefont{and}
  \bibinfo{author}{\bibfnamefont{K.}~\bibnamefont{Takahashi}},
  \bibinfo{journal}{Journal of Crystal Growth} \textbf{\bibinfo{volume}{128}},
  \bibinfo{pages}{550 } (\bibinfo{year}{1993}).

\bibitem[{\citenamefont{Brunemeier}(1991)}]{BrunemeierJVacTechnolB91}
\bibinfo{author}{\bibfnamefont{P.}~\bibnamefont{Brunemeier}},
  \bibinfo{journal}{J.Vac.Sci.Technol.B 9(5)} p. \bibinfo{pages}{2554}
  (\bibinfo{year}{1991}).

\bibitem[{\citenamefont{Herman and Sitter}(1989)}]{Herman89MBE}
\bibinfo{author}{\bibfnamefont{M.}~\bibnamefont{Herman}} \bibnamefont{and}
  \bibinfo{author}{\bibfnamefont{H.}~\bibnamefont{Sitter}},
  \emph{\bibinfo{title}{Molecular Beam Epitaxy}}
  (\bibinfo{publisher}{Springer-Verlag Berlin}, \bibinfo{year}{1989}).

\bibitem[{\citenamefont{Clarke}()}]{ClarkePrivateCommunication}
\bibinfo{author}{\bibfnamefont{E.}~\bibnamefont{Clarke}},
  \bibinfo{note}{private communication}.

\bibitem[{\citenamefont{Langbein}(2010)}]{LangbeinRNC10}
\bibinfo{author}{\bibfnamefont{W.}~\bibnamefont{Langbein}},
  \bibinfo{journal}{Rivista del nuovo cimento} \textbf{\bibinfo{volume}{33}},
  \bibinfo{pages}{255} (\bibinfo{year}{2010}).

\bibitem[{\citenamefont{Giannuzzi and Stevie}(2005)}]{Giannuzzi05FIB}
\bibinfo{editor}{\bibfnamefont{L.~A.} \bibnamefont{Giannuzzi}}
  \bibnamefont{and} \bibinfo{editor}{\bibfnamefont{F.~A.}
  \bibnamefont{Stevie}}, eds., \emph{\bibinfo{title}{Introduction to focused
  ion beams: instrumentation, theory, techniques, and practise}}
  (\bibinfo{publisher}{Springer-Verlag Berlin}, \bibinfo{year}{2005}).

\bibitem[{\citenamefont{Abramowitz and Stegun}(1965)}]{Abramowitz65}
\bibinfo{author}{\bibfnamefont{M.}~\bibnamefont{Abramowitz}} \bibnamefont{and}
  \bibinfo{author}{\bibfnamefont{I.}~\bibnamefont{Stegun}},
  \emph{\bibinfo{title}{Handbook of Mathematical Functions: with Formulas,
  Graphs, and Mathematical Tables}} (\bibinfo{publisher}{Dover Publications},
  \bibinfo{year}{1965}).

\bibitem[{\citenamefont{McLachlan}(1947)}]{McLachlan47}
\bibinfo{author}{\bibfnamefont{N.}~\bibnamefont{McLachlan}},
  \emph{\bibinfo{title}{Theory and applications of Mathieu functions}}
  (\bibinfo{publisher}{Oxford}, \bibinfo{year}{1947}).

\bibitem[{\citenamefont{Kaitouni et~al.}(2006)\citenamefont{Kaitouni, Daïf,
  Baas, Richard, Paraiso, Lugan, Guillet, Morier-Genoud, Ganière, Staehli
  et~al.}}]{KaitouniPRB06}
\bibinfo{author}{\bibfnamefont{R.~I.} \bibnamefont{Kaitouni}},
  \bibinfo{author}{\bibfnamefont{O.~E.} \bibnamefont{Daïf}},
  \bibinfo{author}{\bibfnamefont{A.}~\bibnamefont{Baas}},
  \bibinfo{author}{\bibfnamefont{M.}~\bibnamefont{Richard}},
  \bibinfo{author}{\bibfnamefont{T.}~\bibnamefont{Paraiso}},
  \bibinfo{author}{\bibfnamefont{P.}~\bibnamefont{Lugan}},
  \bibinfo{author}{\bibfnamefont{T.}~\bibnamefont{Guillet}},
  \bibinfo{author}{\bibfnamefont{F.}~\bibnamefont{Morier-Genoud}},
  \bibinfo{author}{\bibfnamefont{J.~D.} \bibnamefont{Ganière}},
  \bibinfo{author}{\bibfnamefont{J.~L.} \bibnamefont{Staehli}},
  \bibnamefont{et~al.}, \bibinfo{journal}{Phys. Rev. B}
  \textbf{\bibinfo{volume}{74}}, \bibinfo{pages}{155311}
  (\bibinfo{year}{2006}).

\bibitem[{\citenamefont{Waalkens et~al.}(1997)\citenamefont{Waalkens, Wiersig,
  and Dullin}}]{WaalkensAnnPhys97}
\bibinfo{author}{\bibfnamefont{H.}~\bibnamefont{Waalkens}},
  \bibinfo{author}{\bibfnamefont{J.}~\bibnamefont{Wiersig}}, \bibnamefont{and}
  \bibinfo{author}{\bibfnamefont{H.~R.} \bibnamefont{Dullin}},
  \bibinfo{journal}{Ann. Phys} \textbf{\bibinfo{volume}{260}}
  (\bibinfo{year}{1997}).

\bibitem[{\citenamefont{Coisson et~al.}(2009)\citenamefont{Coisson, Vernizzi,
  and Yang}}]{Scilab}
\bibinfo{author}{\bibfnamefont{R.}~\bibnamefont{Coisson}},
  \bibinfo{author}{\bibfnamefont{G.}~\bibnamefont{Vernizzi}}, \bibnamefont{and}
  \bibinfo{author}{\bibfnamefont{X.}~\bibnamefont{Yang}}, pp. \bibinfo{pages}{3
  --10} (\bibinfo{year}{2009}).

\bibitem[{\citenamefont{Mano et~al.}(2005)\citenamefont{Mano, Kuroda,
  Sanguinetti, Ochiai, Tateno, Kim, Noda, Kawabe, Sakoda, Kido
  et~al.}}]{ManoNL05}
\bibinfo{author}{\bibfnamefont{T.}~\bibnamefont{Mano}},
  \bibinfo{author}{\bibfnamefont{T.}~\bibnamefont{Kuroda}},
  \bibinfo{author}{\bibfnamefont{S.}~\bibnamefont{Sanguinetti}},
  \bibinfo{author}{\bibfnamefont{T.}~\bibnamefont{Ochiai}},
  \bibinfo{author}{\bibfnamefont{T.}~\bibnamefont{Tateno}},
  \bibinfo{author}{\bibfnamefont{J.}~\bibnamefont{Kim}},
  \bibinfo{author}{\bibfnamefont{T.}~\bibnamefont{Noda}},
  \bibinfo{author}{\bibfnamefont{M.}~\bibnamefont{Kawabe}},
  \bibinfo{author}{\bibfnamefont{K.}~\bibnamefont{Sakoda}},
  \bibinfo{author}{\bibfnamefont{G.}~\bibnamefont{Kido}}, \bibnamefont{et~al.},
  \bibinfo{journal}{Nano Letters} \textbf{\bibinfo{volume}{5}},
  \bibinfo{pages}{425} (\bibinfo{year}{2005}).

\bibitem[{\citenamefont{Koshiba et~al.}(1994)\citenamefont{Koshiba, Nakamura,
  Tsuchiya, Noge, Kano, Nagamune, Noda, and Sakaki}}]{KoshibaAPL94}
\bibinfo{author}{\bibfnamefont{S.}~\bibnamefont{Koshiba}},
  \bibinfo{author}{\bibfnamefont{Y.}~\bibnamefont{Nakamura}},
  \bibinfo{author}{\bibfnamefont{M.}~\bibnamefont{Tsuchiya}},
  \bibinfo{author}{\bibfnamefont{H.}~\bibnamefont{Noge}},
  \bibinfo{author}{\bibfnamefont{H.}~\bibnamefont{Kano}},
  \bibinfo{author}{\bibfnamefont{Y.}~\bibnamefont{Nagamune}},
  \bibinfo{author}{\bibfnamefont{T.}~\bibnamefont{Noda}}, \bibnamefont{and}
  \bibinfo{author}{\bibfnamefont{H.}~\bibnamefont{Sakaki}},
  \bibinfo{journal}{Journal of Applied Physics} \textbf{\bibinfo{volume}{76}},
  \bibinfo{pages}{4138} (\bibinfo{year}{1994}).

\bibitem[{\citenamefont{Ohta et~al.}(1989)\citenamefont{Ohta, Kojima, and
  Nakagawa}}]{OhtaJCG89}
\bibinfo{author}{\bibfnamefont{K.}~\bibnamefont{Ohta}},
  \bibinfo{author}{\bibfnamefont{T.}~\bibnamefont{Kojima}}, \bibnamefont{and}
  \bibinfo{author}{\bibfnamefont{T.}~\bibnamefont{Nakagawa}},
  \bibinfo{journal}{Journal of Crystal Growth} \textbf{\bibinfo{volume}{95}},
  \bibinfo{pages}{71 } (\bibinfo{year}{1989}), ISSN \bibinfo{issn}{0022-0248}.

\bibitem[{\citenamefont{Biegelsen et~al.}(1990)\citenamefont{Biegelsen,
  Bringans, Northrup, and Swartz}}]{BiegelsenPRB90}
\bibinfo{author}{\bibfnamefont{D.~K.} \bibnamefont{Biegelsen}},
  \bibinfo{author}{\bibfnamefont{R.~D.} \bibnamefont{Bringans}},
  \bibinfo{author}{\bibfnamefont{J.~E.} \bibnamefont{Northrup}},
  \bibnamefont{and} \bibinfo{author}{\bibfnamefont{L.-E.}
  \bibnamefont{Swartz}}, \bibinfo{journal}{Phys. Rev. B}
  \textbf{\bibinfo{volume}{41}}, \bibinfo{pages}{5701} (\bibinfo{year}{1990}).

\bibitem[{\citenamefont{Liew et~al.}(2011)\citenamefont{Liew, Shelykh, and
  Malpuech}}]{LiewPhysicaE11}
\bibinfo{author}{\bibfnamefont{T.}~\bibnamefont{Liew}},
  \bibinfo{author}{\bibfnamefont{I.}~\bibnamefont{Shelykh}}, \bibnamefont{and}
  \bibinfo{author}{\bibfnamefont{G.}~\bibnamefont{Malpuech}},
  \bibinfo{journal}{Physica E: Low-dimensional Systems and Nanostructures}
  \textbf{\bibinfo{volume}{43}}, \bibinfo{pages}{1543 } (\bibinfo{year}{2011}).

\bibitem[{\citenamefont{Cerna et~al.}(2009)\citenamefont{Cerna, Sarchi,
  Paraïso, Nardin, L\`eger, Richard, Pietka, Daif, Morier-Genoud, Savona
  et~al.}}]{CernaPRB09}
\bibinfo{author}{\bibfnamefont{R.}~\bibnamefont{Cerna}},
  \bibinfo{author}{\bibfnamefont{D.}~\bibnamefont{Sarchi}},
  \bibinfo{author}{\bibfnamefont{T.~K.} \bibnamefont{Paraïso}},
  \bibinfo{author}{\bibfnamefont{G.}~\bibnamefont{Nardin}},
  \bibinfo{author}{\bibfnamefont{Y.}~\bibnamefont{L\`eger}},
  \bibinfo{author}{\bibfnamefont{M.}~\bibnamefont{Richard}},
  \bibinfo{author}{\bibfnamefont{B.}~\bibnamefont{Pietka}},
  \bibinfo{author}{\bibfnamefont{O.~E.} \bibnamefont{Daif}},
  \bibinfo{author}{\bibfnamefont{F.}~\bibnamefont{Morier-Genoud}},
  \bibinfo{author}{\bibfnamefont{V.}~\bibnamefont{Savona}},
  \bibnamefont{et~al.}, \bibinfo{journal}{Phys. Rev. B}
  \textbf{\bibinfo{volume}{80}}, \bibinfo{pages}{121309(R)}
  (\bibinfo{year}{2009}).

\bibitem[{\citenamefont{Mantovani et~al.}(2004)\citenamefont{Mantovani,
  Sanguinetti, Guzzi, Grilli, Gurioli, Watanabe, and Koguchi}}]{MantovaniJAP04}
\bibinfo{author}{\bibfnamefont{V.}~\bibnamefont{Mantovani}},
  \bibinfo{author}{\bibfnamefont{S.}~\bibnamefont{Sanguinetti}},
  \bibinfo{author}{\bibfnamefont{M.}~\bibnamefont{Guzzi}},
  \bibinfo{author}{\bibfnamefont{E.}~\bibnamefont{Grilli}},
  \bibinfo{author}{\bibfnamefont{M.}~\bibnamefont{Gurioli}},
  \bibinfo{author}{\bibfnamefont{K.}~\bibnamefont{Watanabe}}, \bibnamefont{and}
  \bibinfo{author}{\bibfnamefont{N.}~\bibnamefont{Koguchi}},
  \bibinfo{journal}{Journal of Applied Physics} \textbf{\bibinfo{volume}{96}},
  \bibinfo{pages}{4416} (\bibinfo{year}{2004}).

\bibitem[{\citenamefont{Paraiso et~al.}(2010)\citenamefont{Paraiso, Wouters,
  Leger, Morier-Genoud, and Deveaud-Plédran}}]{ParaisoNatMatt10}
\bibinfo{author}{\bibfnamefont{T.}~\bibnamefont{Paraiso}},
  \bibinfo{author}{\bibfnamefont{M.}~\bibnamefont{Wouters}},
  \bibinfo{author}{\bibfnamefont{Y.}~\bibnamefont{Leger}},
  \bibinfo{author}{\bibfnamefont{F.}~\bibnamefont{Morier-Genoud}},
  \bibnamefont{and}
  \bibinfo{author}{\bibfnamefont{B.}~\bibnamefont{Deveaud-Plédran}},
  \bibinfo{journal}{Nature Materials} \textbf{\bibinfo{volume}{9}},
  \bibinfo{pages}{655?660} (\bibinfo{year}{2010}).

\end{thebibliography}

\end{document}